\documentclass[aps,pra,amsmath,amssymb,amsfonts,a4paper,10pt,nofootinbib,twocolumn]{revtex4-2}  
\usepackage{mathrsfs}
\usepackage{verbatim}
\usepackage{epsfig}
\usepackage{color}
\usepackage{soul}
\usepackage[english]{babel}
\usepackage[normalem]{ulem}
\usepackage{orcidlink}
\usepackage{graphicx}
\usepackage{derivative}
\usepackage{subcaption}
\usepackage{enumitem} 
\usepackage{dsfont}
\usepackage{siunitx}
\usepackage{float} 

\newcommand*{\bra}[1]{\ensuremath{\langle{#1}\rvert}}
\newcommand*{\ket}[1]{\ensuremath{\lvert{#1}\rangle}}

\begin{document}

\title{Optomechanical Systems with Linear and Quadratic Position Couplings: \\
Dynamics and Optimal Estimation}
\author{Yaqing Xy Wang}
\email{yaq.wang@fz-juelich.de}
\affiliation{Institute for Theoretical Physics, University of Cologne, 50937 Köln, Germany}
\affiliation{Forschungszentrum Jülich, Institute of Quantum Control,
Peter Grünberg Institut (PGI-8), 52425 Jülich, Germany}

\author{Claudio Sanavio}
\email{claudio.sanavio@iit.it}
\affiliation{Fondazione Istituto Italiano di Tecnologia, 
Center for Life Nano-Neuroscience at la Sapienza, 
Viale Regina Elena 291, 00161 Roma, Italy}

\author{J\'ozsef Zsolt Bern\'ad}
\email{j.bernad@fz-juelich.de}
\affiliation{Forschungszentrum Jülich, Institute of Quantum Control,
Peter Grünberg Institut (PGI-8), 52425 Jülich, Germany}
\affiliation{HUN-REN Wigner Research Centre for Physics, Budapest, Hungary}

\begin{abstract}
We study the dynamics of an optomechanical system consisting of a single-mode optical field coupled to a mechanical oscillator, where the nonlinear interaction includes both linear and quadratic terms in the oscillator’s position. We present an analytical solution to this quantum mechanical Hamiltonian problem by employing the formalism of two-phonon coherent states. Quantum estimation theory is applied to the resulting state of the optical field, with a focus on evaluating the quantum Fisher information with respect to the strength of the quadratic coupling. Our estimation scheme employs balanced homodyne photodetection and demonstrates that the corresponding classical Fisher information can reach the quantum Fisher information limit, with the phase of the local coherent oscillator playing a crucial role. 
\end{abstract}

\maketitle

\section{Introduction}
\label{sec:introduction}

Optomechanical systems~\cite{aspelmeyer_cavity_2014} are at the center of a technological revolution~\cite{barzanjeh_optomechanics_2022} that is based on the interaction between light and matter to study fundamental physics~\cite{qvarfort_gravimetry_2018} and produce innovative devices, 
spanning from signal processors~\cite{andrews_quantum-enabled_2015} to transductors~\cite{forsch_microwave--optics_2020} and quantum amplifiers~\cite{peano_topological_2016,sanavio_nonreciprocal_2020}. 
Achieving these goals requires a thorough understanding of the optomechanical system, which can be attained through precise measurements of its parameters.
Numerous studies~\cite{bernad_optimal_2018,sanavio_fisher-information-based_2020,schneiter_optimal_2020,carrasco_estimation_2022,ruppert_high-precision_2022,sanavio_estimation_2022, Peng:24, Peng:25} have focused on understanding how to estimate the parameters of an optical cavity coupled to a mechanical oscillator via radiation pressure.

In constructing effective estimation frameworks, the first essential step is to define a statistical model for the data \cite{vanTrees, Kay}. In quantum mechanics, this requirement translates to obtaining an exact solution for the time evolution of the quantum state. Formally, this yields a family of quantum states parameterized by the unknown variables to be estimated. However, due to the complexity introduced by the high-dimensional parameter spaces commonly encountered in optomechanical systems, it is often necessary to introduce simplifications to the physical model. 
These approximations facilitate the formulation of a minimal model that remains tractable while still capturing the essential physics. A detailed derivation of the optomechanical Hamiltonian is presented in the foundational works \cite{Law94, law_interaction_1995}. Two principal approximations are typically employed to simplify this Hamiltonian. First, if the mechanical oscillator evolves adiabatically slowly compared to the frequency separation between optical modes, intermodal photon scattering can be neglected. Under this condition, the system can be effectively described using a single optical field mode. Second, in the linear approximation, the radiation–mechanical interaction is simplified by expanding the coupling Hamiltonian to first order in the mechanical position operator. The first objective of this paper is to construct a statistical model by deriving the exact time-evolving quantum state governed by the optomechanical Hamiltonian, under the adiabatic approximation while extending the standard linear interaction to include second-order (quadratic) terms in the mechanical displacement.

The Hamiltonian under investigation has been previously discussed in comprehensive reviews \cite{aspelmeyer_cavity_2014}. It has found applications in various contexts, including quantum nondemolition measurements of the phonon number of the mechanical mode \cite{Milburn}, as well as in optomechanically induced parametric oscillations \cite{Fainstein}. Mathematical analyses of such systems often concentrate on the corresponding Heisenberg equations of motion and their expectation values. In many cases, these approaches rely on suitable approximations to obtain a finite set of coupled differential equations describing the system dynamics \cite{Khorasani}. An alternative strategy is to confine the dynamics to Gaussian states, which enables the use of general closed form expressions in quantum parameter estimation studies \cite{Sala}. The aforementioned approaches can even be extended to include dissipative processes within the Hamiltonian framework. In contrast, the present work is concerned with obtaining an exact analytical solution for a general time-dependent quantum state, without imposing Gaussian or perturbative approximations. To the best of our knowledge, such an exact formulation has not been reported previously. We employ the formalism of two-photon coherent states \cite{yuen_two-photon_1976}, adapted here to describe the mechanical oscillator. The obtained solutions describe the joint quantum state of the optical field and the mechanical oscillator. Since, in practice, measurements are typically performed on the optical field, we trace out the mechanical degrees of freedom. We provide a comprehensive analysis on the resulting state and its dynamics.

In the quantum parameter estimation analysis, we employ a frequentist approach to determine the lower bounds on the variance of any unbiased estimator of the quadratic optomechanical coupling. Even without explicitly constructing the estimators, these bounds serve as benchmark values for assessing the performance of those used in experiments. In this context, we analyze the quantum Fisher information (QFI), which is inversely proportional to the lower bound in the Cramér-Rao inequality~\cite{helstrom_quantum_1969, Paris}. We determine the QFI for the optical field state as a function of the unknown strength of the quadratic optomechanical coupling. This is compared with the classical Fisher information (CFI), obtained for balanced-homodyne photodetection measurement scheme. The QFI serves as a measure of the sensitivity of the optical field state to changes in the quadratic optomechanical coupling constant; in other words, it quantifies how much information the quantum state contains about this parameter. However, the CFI assesses the potential to evaluate this sensitivity using a set of classical measurements.


This paper is organized as follows. In Sec.~\ref{sec:model}, we describe the analytical model and we solve the equations of motion, with explicit dependence upon the linear and quadratic (the mechanical position operator) optomechanical coupling constants. In Sec.~\ref{sec:III} we introduce the quantum and classical Fisher information and the projective measures describing the standard and the  balanced-homodyne photodetection schemes.
In Sec.~\ref{sec:IV} we show the dynamics of the optical state affected by quadratic optomechanical coupling, and we evaluate the quantum and classical Fisher information, while analyzing the performance of the two measurement schemes. Finally, in Sec.~\ref{sec:V} we draw our conclusions and outlooks.

\section{Model}
\label{sec:model}

We consider a single-mode Fabry-P\'erot optical cavity with a suspended dielectric membrane that is free to oscillate. The electromagnetic field inside the cavity exerts radiation pressure on the mechanical oscillator and causes it to oscillate around its equilibrium position. The Hamiltonian of the optomechanical system can be written by summing the contribution from the optical and the mechanical subsystems as~\cite{law_interaction_1995}
\begin{equation}
\label{eq:1}
\hat{H} = \frac{\hat{p}^2}{2m}+\frac{1}{2}m\Omega^2\hat{x}^2+\hbar\omega(\hat{x})\hat{a}^\dagger\hat{a},
\end{equation}
where $\hat{x}$ is the position operator of the mechanical system, $\hat{p}$ is its momentum operator, $m$ is its mass, and $\Omega$ is the mechanical frequency of motion. 
The operators $\hat{a},\hat{a}^{\dagger}$ are the photonic ladder operators with the commutation relation $[\hat{a},\hat{a}^{\dagger}]=1$, and $\omega(x)$ is the resonant optical frequency of the cavity, which depends on the position $x$ of the mechanical oscillator. 
If we expand the optical frequency $\omega(x)$ around the equilibrium value $x_0$, we obtain
\begin{eqnarray}\label{eq:expansion}
\omega(x) &\approx& \omega(x_0) +\omega'(x_0)(x-x_0)+\frac{\omega''(x_0)}{2!}(x-x_0)^2 \nonumber \\
&=&\omega_c +g_1 x + g_2 x^2,
\end{eqnarray}
where $\omega^{(n)}(x_0)$ denotes the $n$th derivative of $\omega(x)$ evaluated at the point $x_0$. Here, $g_1$ is the linear, and $g_2$ is the quadratic optomechanical coupling constant. The parameter $\omega_c$ is identified as the frequency of the single mode of the radiation field. It is worth noting that by combining Eqs. \eqref{eq:1} and \eqref{eq:expansion}, the harmonic potential governing the mechanical system takes on a modified form: $m\Omega^2\hat{x}^2/2 + \hbar g_2 \hat{a}^\dagger\hat{a} \hat{x}^2 $, where the position is treated as an operator. To preserve the harmonic motion of the mechanical system, we must ensure that for any physically realizable quantum state $\ket{\Psi}_c$ of the cavity field, the coefficient of $\hat{x}^2$
remains positive; that is,
\begin{equation}
 m\Omega^2/2+\hbar g_2 \,{}_c\bra{\Psi} \hat{a}^\dagger\hat{a} \ket{\Psi}_c>0. 
 \label{eq:stability}
\end{equation}
This condition not only guarantees harmonic behavior but also serves as a stability criterion for the model, ensuring the system consists of two interacting quantum harmonic oscillators. Under this constraint, the optomechanical Hamiltonian becomes:
\begin{equation}\label{eq:Hamiltonian}
\hat{H}^{\text{quad}} = \frac{\hat{p}^2}{2m}+\frac{1}{2}m\Omega^2\hat{x}^2+\hbar\omega_c\hat{a}^\dagger\hat{a}+\hbar g_1\hat{a}^\dagger\hat{a}\hat{x}+ \hbar g_2\hat{a}^\dagger\hat{a}\hat{x}^2.
\end{equation}
It is worth mentioning that the single-mode approach requires the motion of the mechanical oscillator to be adiabatically slow, so the model can ignore the scattering of photons from the mode to other cavity modes~\cite{Law94}. Furthermore, this unitary model assumes that the cavity is lossless and the mechanical oscillator is not subject to any decoherence processes.

In order to solve the dynamics, we notice that the Hamiltonian~\eqref{eq:Hamiltonian} is block diagonal in the photonic Fock basis $|n\rangle_c$ ($n\in {\mathbb N}_0$), i.e. it can be written as
\begin{equation}
\hat{H}^{\text{quad}}= \sum_n \hat{H}_n, \quad \hat{H}_n=\hat{P}_n\hat{H}^{\text{quad}}\hat{P}_n,
\end{equation}
where $\hat{P}_n=|n\rangle_c\langle n|$ is the projector on the subspace with $n$ photons. 
The Hamiltonian $\hat{H}_n$ reads
\begin{equation}\label{eq:Hamiltonian_nl}
\hat{H}_n=\hbar n\omega_c \hat{I}+\frac{\hat{p}^2}{2m}+\frac{1}{2}m\bigg{(}\Omega^2+\frac{2\hbar g_2n}{m}\bigg{)}\hat{x}^2+\hbar g_1n\hat{x},
\end{equation}
where $\hat{I}$ is the identity operator on the Hilbert space of the mechanical oscillator. By setting 
\begin{equation}
\Omega_n=\sqrt{\Omega^2+\frac{2\hbar g_2n}{m}},  \label{eq:Omn}  
\end{equation}
which due to \eqref{eq:stability} is always strictly positive, we can define the ladder mechanical operators $\hat{b}_n,\hat{b}^{\dagger}_n$ as
\begin{eqnarray}\label{eq:b_n}
\hat{b}_n&=&\sqrt{\frac{m\Omega_n}{2\hbar }}\bigg{(}\hat{x}+i\frac{\hat{p}}{m\Omega_n}\bigg{)},\\
\hat{b}^{\dagger}_n&=&\sqrt{\frac{m\Omega_n}{2\hbar }}\bigg{(}\hat{x}-i\frac{\hat{p}}{m\Omega_n}\bigg{)}.
\end{eqnarray}
We stress that the ladder operators related to different indices $\hat{b}_n,\hat{b}_l$ with $n\neq l$ do not commute. In fact, the relations between them are given by the following Bogoliubov-like transformation rules:
\begin{eqnarray}\label{eq:Bogoliubov}
\hat{b}_l&=&\frac{1}{2}\sqrt{\frac{\Omega_l}{\Omega_n}}\bigg{[}\bigg{(}1-\frac{\Omega_n}{\Omega_l}\bigg{)}\hat{b}_n^\dagger+\bigg{(}1+\frac{\Omega_n}{\Omega_l}\bigg{)}\hat{b}_n\bigg{]},\nonumber\\
\hat{b}_l^\dagger&=&\frac{1}{2}\sqrt{\frac{\Omega_l}{\Omega_n}}\bigg{[}\bigg{(}1-\frac{\Omega_n}{\Omega_l}\bigg{)}\hat{b}_n+\bigg{(}1+\frac{\Omega_n}{\Omega_l}\bigg{)}\hat{b}_n^\dagger\bigg{]}.
\end{eqnarray}
This must be taken into account when solving the dynamics of the system, because any initial state of the mechanical oscillator is represented by the Fock basis $|m\rangle_0$ ($m\in {\mathbb N}_0$) defined by the ladder operators $\hat{b}_0,\hat{b}^{\dagger}_0$. The Hamiltonian~\eqref{eq:Hamiltonian_nl} assumes the form
\begin{equation}
\label{eq:nHamiltonian}
\hat{H}_n=\hbar \omega_n\hat{I}+\hbar \Omega_n\hat{b}^{\dagger}_n\hat{b}_n+ \hbar g_{1,n}(\hat{b}_n+\hat{b}^{\dagger}_n),
\end{equation}
where 
\begin{equation}
\omega_n=n\omega_c+\Omega_n/2 \label{eq:omn}   
\end{equation} and $g_{1,n}=ng_1\sqrt{\frac{\hbar }{2m\Omega_n}}$. By applying the Baker-Campbell-Hausdorff formula, we get~\cite{bernad_centre--mass_2013}
\begin{equation}\label{eq:nonlinear_sol_n}
e^{-i\hat{H}_nt/\hbar }= e^{-i\Phi_n(t)}e^{\eta_n(t)\hat{b}_n^{\dagger}-\eta_n^*(t)\hat{b}_n}e^{-i\Omega_n\hat{b}^\dagger_n\hat{b}_nt},
\end{equation}
where we have introduced the functions
\begin{eqnarray}
\label{eq:Phi_n}
\Phi_n(t)&=&\omega_nt-\frac{g^2_{1,n}}{\Omega_n^2}[\Omega_n t-\sin(\Omega_n t)],\\ \label{eq:beta_n}
\eta_n(t)&=&\frac{g_{1,n}}{\Omega_n}(e^{-i\Omega_n t}-1).
\end{eqnarray}
Given an initial state $\hat{\varrho}(0)$, the time evolution of the system is given by the von Neumann equation
\begin{equation}
\label{eq:von Neumann}
 \hat{\varrho}(t)=e^{-i \hat{H}^{\text{quad}} t/\hbar } \hat{\varrho}(0) e^{i \hat{H}^{\text{quad}} t/\hbar }.
\end{equation}
We are interested in cases where no initial correlations between the single-mode optical field and the mechanical oscillator are considered. Therefore, we choose an initial state of the form
\begin{eqnarray}\label{eq:rho_0}
\hat{\varrho}(0) = \hat{\rho}(0) \otimes \mathfrak{\hat{m}}(0),
\end{eqnarray}
where $\hat \rho$ and $\mathfrak{\hat{m}}$ are the optical and the mechanical density operator, respectively.

Now, we consider the density operator of the mechanical oscillator in the following representation:
\begin{equation}\label{eq:initial_mechanical_state}
\mathfrak{\hat{m}}(0)=\int d^2\alpha \, P(\alpha,\alpha^*)|\alpha\rangle_0\langle \alpha |.    
\end{equation}
where $d^2 \alpha=d\mathrm{Re}\{\alpha\}\,d\mathrm{Im}\{\alpha\}$, $\ket{\alpha}_0$ is a coherent state, and $P(\alpha,\alpha^*)$ is a Glauber-Sudarshan phase-space distribution \cite{Glauber,Sudarshan}. Due to the fact that $\hat{H}^{\text{quad}}$ is block diagonal in the photonic Fock basis, we choose
\begin{equation}
 \hat{\rho}(0)=\sum^\infty_{n,m=0} a_{n,m} \ket{n}_c\bra{m},
\end{equation}
where the matrix entries $a_{n,m}$ are subject to the conditions
\begin{equation}
    \mathrm{Tr} \{\hat{\rho}(0)\}=1 \quad \text{and} \quad \hat{\rho}(0) \geq 0,
\end{equation}
i.e., $\hat{\rho}(0)$ is a positive semidefinite operator with trace one.

To evaluate the time evolution of the state, we employ the Bogoliubov transformations~\eqref{eq:Bogoliubov} to calculate the action of $e^{-iH_nt/\hbar }$ on the eigenstates of the number operator $\hat{N}_n = \hat{b}^\dagger_n\hat{b}_n$. They define a Fock basis $|m\rangle_n$ ($m\in {\mathbb N}_0$), whose properties have been extensively investigated in the literature~\cite{yuen_two-photon_1976,yuen_optical_1978,shapiro_optical_1979,mandel_optical_1995}. The displacement operator $\hat{D}_n(\alpha) = e^{\alpha\hat{b}^{\dagger}_n-\alpha^*\hat{b}_n}$ has the following property
\begin{eqnarray}\label{eq:two-photon-coherent}
    \hat{D}_n(\alpha)|0\rangle_n =|\alpha\rangle_n,
\end{eqnarray}
where $\ket{\alpha}_n$ is an example of a two-phonon coherent state. We note that the concept of two-photon coherent states was originally developed for quantized electromagnetic fields ~\cite{yuen_two-photon_1976}; however, in this work, we apply the formalism to an  oscillator with mass. With the help of the unitary squeeze operator
\begin{equation}
   \hat{S}(z) = e^{ \frac{1}{2} (z^* \hat{b}^2_0 - z \hat{b}^{\dagger 2}_0)}, \quad z = r \, e^{i\theta},
\end{equation}
we have the transformation 
\begin{eqnarray}
    \ket{m}_n&=&\hat{S}(z_n) \ket{m}_0, \quad z_n = -r_n \, e^{i\theta_n}, \\
    \cosh r_n &=& \frac{1}{2}\sqrt{\frac{\Omega_n}{\Omega}}\bigg{(}1+\frac{\Omega}{\Omega_n}\bigg{)}, 
    \quad \theta_n=0.
\end{eqnarray}
Now, we can evaluate Eq.~\eqref{eq:von Neumann} as 
\begin{widetext}
\begin{eqnarray}
\hat{\varrho}(t) &=&\sum^\infty_{n,m=0} a_{n,m} \ket{n}_c\bra{m}e^{-i\hat{H}_nt/\hbar }\hat{\mathfrak{m}}(0)e^{i\hat{H}_mt/\hbar }\nonumber\\
&=&\sum^\infty_{n,m=0} a_{n,m} \ket{n}_c\bra{m}e^{-i\left[\Phi_n(t)-\Phi_m(t)\right]}\int d^2\alpha\, P(\alpha,\alpha^*)D_n\left[\eta_n(t)\right]e^{-i\Omega_n\hat{N}_nt}|\alpha\rangle_0\langle\alpha|e^{i\Omega_m\hat{N}_mt}D^{\dagger}_m\left[\eta_m(t)\right].
\end{eqnarray}
In the next step, we introduce two identity operators
\begin{equation}
    \hat{I}=\frac{1}{\pi} \int \ket{\beta}_n\bra{\beta}\, d^{2}\beta=\frac{1}{\pi} \int \ket{\gamma}_m\bra{\gamma}\, d^{2}\gamma
\end{equation}
to get
\begin{eqnarray}
\hat{\varrho}(t)&=& \sum^\infty_{n,m=0} a_{n,m} \ket{n}_c\bra{m}e^{-i\left[\Phi_n(t)-\Phi_m(t)\right]}  \nonumber \\
&\times& \int \frac{P(\alpha,\alpha^*)}{\pi^2} \hat{D}_n\left[\eta_n(t)\right]e^{-i\Omega_n\hat{b}^\dagger_n \hat{b}_nt}\ket{\beta}_n\langle \beta \ket{\alpha}_0\bra{\alpha} \gamma \rangle_m\bra{\gamma} e^{i\Omega_m\hat{b}^\dagger_m \hat{b}_mt}\hat{D}^{\dagger}_m\left[\eta_m(t)\right]\,d^{2}\alpha \, d^{2}\beta \, d^{2}\gamma \\
&=& \sum^\infty_{n,m=0} a_{n,m} \ket{n}_c\bra{m}e^{-i\left[\Phi_n(t)-\Phi_m(t)\right]} \int \frac{P(\alpha,\alpha^*)}{\pi^2} \ket{\eta_n(t)+\beta e^{-i\Omega_n t}}_n\langle\beta
 \ket{\alpha}_0 \bra{\alpha} \gamma \rangle_m
  \bra{ \eta_m(t)+\gamma e^{-i\Omega_mt}}  \nonumber \\
  &\times& e^{\frac{1}{2} 
    \left[\eta_n(t) \beta^* e^{i \Omega_n t} - \eta^*_n(t)\beta e^{-i \Omega_n t}\right]}
    e^{\frac{1}{2} \left[ \eta_m^*(t) \gamma e^{-i \Omega_m t} - \eta_m(t)\gamma^* e^{i \Omega_m t}\right]}\, d^{2}\alpha \, d^{2}\beta \, d^{2}\gamma,
\end{eqnarray}
\end{widetext}
 where we have used the relations:
\begin{eqnarray}
    e^{-i\Omega \hat{b}^\dagger_n \hat{b}_nt}|\alpha\rangle_n &= &|\alpha e^{-i \Omega t}\rangle_n,\\
    \hat{D}_n(\beta)|\alpha\rangle_n &=& 
    e^{\frac{1}{2}(\beta \alpha^* - \beta^*\alpha)}
    |\beta+\alpha\rangle_n.
\end{eqnarray}
The remainder of the calculation will focus solely on the optical density operator, $\hat \rho(t)$, at time $t$. This approach is justified by the fact that, in most practical scenarios, the optical field is the only component of the system that is experimentally accessible. The operator $\hat \rho(t)$ is obtained by performing a partial trace over the mechanical degrees of freedom, and it can be expressed as
\begin{equation}
 \hat{\rho}(t)=\mathrm{Tr}_{\text{mech}}\{\hat{\varrho}(t)\}   = \sum^\infty_{n,m=0}a_{n,m}(t)|n\rangle_c\langle m|,
\end{equation}
where
\begin{widetext}
\begin{equation}\label{eq:optical_coefficient_1}
a_{n,m}(t) = a_{n,m}e^{-i\left[\Phi_n(t)-\Phi_m(t)\right]}  \int \frac{P(\alpha,\alpha^*)}{\pi^3} {}_n\langle\beta
 \ket{\alpha}_0 \bra{\alpha} \gamma \rangle_m \,{}_0\langle \delta
 \ket{\eta_n(t)+\beta e^{-i\Omega_n t}}_n \, {}_m\bra{ \eta_m(t)+\gamma e^{-i\Omega_mt}} \delta \rangle_0 \, d^{2}\alpha \, d^{2}\beta \, d^{2}\gamma \, d^{2}\delta
\end{equation}
\end{widetext}
To derive this equation and to subsequently make use of Eq. \eqref{eq:A1}, we have introduced the identity operator
\begin{equation}\label{eq:identity_1}
 \hat{I}=\frac{1}{\pi} \int \ket{\delta}_0\bra{\delta}\, d^{2}\delta.   
\end{equation}
The integral can be evaluated explicitly under the assumption that the mechanical system initially is in a thermal state:
\begin{equation}
    P(\alpha,\alpha^*) = \frac{e^{-|\alpha|^2/n_{\text{th}}}}{\pi n_{\text{th}}}
\end{equation}
with the average phonon number
\begin{equation}\label{eq:thermal_number}
 n_{\text{th}} = \biggl[\exp \biggl(\frac{\hbar\Omega}{k_\text{B}T} \biggr)-1 \biggr]^{-1},
\end{equation}
where $k_\text{B}$ is the Boltzmann constant and $T$ the temperature of the initial thermal state. The expression for $a_{n,m}(t)$ reduces to a Gaussian integral, yielding
\begin{eqnarray}\label{eq:optical_coefficient}
a_{n,m}(t) &=& I\int_{\mathbb{R}_8} \exp (-\frac{1}{2}\textbf{x}^TA\textbf{x} + b^T\textbf{x} + c) d^8\textbf{x} \nonumber \\ 
&=&I \sqrt{\frac{(2\pi)^8}{\textbf{det}(A)}}\exp(\frac{1}{2}b^TA^{-1}b + c),
\end{eqnarray}
where $I$ reads 
\begin{equation}
I =\frac{ a_{n,m} e^{-i\left[\Phi_n(t)-\Phi_m(t)\right]}}{\pi^4 n_{\text{th}} \frac{\Omega_n+\Omega}{2\sqrt{\Omega\Omega_n}} \frac{\Omega_m+\Omega}{2\sqrt{\Omega\Omega_m}} }. 
\end{equation}
The complete expressions for $A$, $b$ and $c$ are provided in Appendix~\ref{app:II}, all derived from the overlaps of two-phonon coherent states~\cite{mandel_optical_1995}, as explicitly given in Eq.~\eqref{eq:A1}. As a consistency check, the initial state in Eq. \eqref{eq:rho_0} was numerically propagated under $\hat{U} = e^{-i\hat{H}^{\textbf{quad}}t}$(see Eq. \eqref{eq:Hamiltonian}) for representative scenarios involving simple parameter values and initial states represented by small size matrices. The mechanical degrees of freedom were then traced out, and agreement with the analytical results in Eq. \eqref{eq:optical_coefficient} was observed within numerical accuracy.

The coefficients $a_{n,m}(t)$ of the optical density operator provide a complete description of the interaction between the single mode of the radiation field and the single vibrational mode of the mechanical oscillator, under the assumption that all losses and sources of decoherence are neglected. We now use these results to estimate the strength of the quadratic optomechanical coupling.

\section{Quantum and Classical Fisher information}\label{sec:III}

In an estimation problem, the goal is to infer the value of an unknown parameter by inspecting datasets
coming from the measurement. In the literature, various approaches to this problem are presented, which broadly fall into two main categories: the Bayesian and frequentist frameworks~\cite{vanTrees,Kay}. In this work, we adopt the frequentist perspective, wherein the parameter is assumed to have a true, fixed (i.e., nonrandom) value. The central question then becomes: what is this true value, and what is the optimal strategy to estimate it? In quantum metrology, this question is often addressed by comparing the classical Fisher information (CFI) with the quantum Fisher information (QFI). 

Given a certain measurement procedure, the CFI quantifies the information that can be extracted from a given state and it is upper-bounded by the QFI, which is the maximization of the CFI over all quantum measurements. Thus, QFI is a property of the state, whereas CFI is a property of both the state and the measurement procedure. Their importance comes from the Cram\'er-Rao theorem \cite{vanTrees} and its quantum mechanical version \cite{helstrom_quantum_1969}, which states the relation between the Fisher information and the minimum attainable variance of any unbiased estimator. It reads

\begin{equation}\label{eq:Cramer_Rao}
\text{Var}_g \geq \frac{1}{N_{\text{meas}} \text{CFI}}\geq \frac{1}{N_{\text{meas}} \text{QFI}},
\end{equation}
where $g$ is the parameter to be estimated.

While the construction of estimators will not be addressed in this work, it plays a crucial role in estimation theory and can be particularly challenging when dealing with probability distributions arising from quantum systems \cite{CkBer}. To be more specific, the measurements that we perform on the system return a set of data, which are then fed to an estimator, a function of the data whose outcome is an estimate of the unknown parameter. Whether the variance of our estimate reaches the minimum given by the inverse of the CFI, following the Cramér-Rao theorem~\eqref{eq:Cramer_Rao}, depends on the estimator itself. For example, the maximum likelihood estimator is able to saturate the lower bound given by the CFI for asymptotically large data sizes~\cite{vanTrees,Kay}.
We point out to the reader that there exist other, less well known versions of the Cramér-Rao theorem for biased estimators~\cite{bernad_optimal_2018}, or when adopting the Bayesian approach~\cite{tsang_ziv-zakai_2012}.

In quantum mechanics, the measurement procedure is described by a positive operator-valued measure (POVM), a collection of operators $\{\hat{\Pi}_i\}_i$, such that any $\hat{\Pi}_i\geq0$ and $\sum_i\hat{\Pi}_i=\hat{I}$, where $\hat{I}$ is the identity operator on the corresponding Hilbert space. We denote the parameter $g$-dependent optical density operator as $\hat{\rho}_g$. The CFI of the quantum state $\hat{\rho}_g$ together with the measurements $\hat{\Pi}_i$ can be calculated as

\begin{equation}\label{eq:CFI}
\text{CFI}=\sum_iP_i(g)\bigg{[}\frac{\partial\ln P_i(g)}{\partial g}\bigg{]}^2,
\end{equation}

\noindent where $P_i(g)=\mathrm{Tr}\{\hat{\rho}_g \hat{\Pi}_i\}$ is the probability of getting the outcome $i$ when the true value of the parameter is $g$.

The QFI could be obtained by maximizing Eq.~\eqref{eq:CFI} over all the possible POVMs, but this is generally a mission impossible. Fortunately, there is a simple expression of it~\cite{helstrom_quantum_1969}:

\begin{equation}\label{eq:QFI}
\text{QFI}=\mathrm{Tr}\{\hat{\rho}_g\hat{L}_g^2\}
\end{equation}

\noindent which is written in terms of the symmetric logarithmic derivative (SLD) $\hat{L}_g$~\cite{holevo_statistical_1973}, which satisfies the relation

\begin{equation}\label{eq:Lyapunov}
\frac{\partial \hat{\rho}_g}{\partial g} = \frac{1}{2}(\hat{L}_g\hat{\rho}_g+\hat{\rho}_g \hat{L}_g).
\end{equation}
The SLD can be found by solving the Lyapunov equation~\eqref{eq:Lyapunov}, with the solution
\begin{equation}
    \hat{L}_g = 2\int_0^\infty e^{-\hat{\rho}_g x} (\partial_g \hat{\rho}_g)e^{-\hat{\rho}_g x} \, dx. 
\end{equation}

We mention here that when the density operator has the property that $\hat{\rho}_g^2=\hat{\rho}_g-h\hat{I}$, with $h\in \mathbb{R}$, which always happens when $\hat{\rho}_g$ describes a two-level system, the SLD assumes the simplified form~\cite{liu_quantum_2016}

\begin{equation}\label{eq:2x2_QFI}
    \hat{L}_g = 2\partial_g\hat{\rho}_g -\frac{1}{2}\partial_g\mathcal{P}\hat{\rho}_g^{-1},
\end{equation}

\noindent where $\hat{\rho}$ is not a pure state, $\mathcal{P}=\mathrm{Tr}\{\hat{\rho}_g^2\}$ is the purity of the system, and it is related to $h$ by
\begin{equation}
\mathcal{P} = 2h +1. 
\end{equation}
This relation can be used to ease the computation of the QFI. In the following we are going to apply Eq.~\eqref{eq:2x2_QFI} to calculate the QFI for the case when the cavity is populated by a superposition of zero-photon and one-photon states. 

In the case of a $2\times2$ system, we use the balanced homodyne photodetection (BHD) measurements in terms of their projective value measures (PVM). The BHD~\cite{tyc_operational_2004} enables the measurement of the observable 
\begin{equation}
    \hat X_\phi = \frac{\hat{a}e^{i\phi}+\hat{a}^\dagger e^{-i\phi}}{2}
\end{equation}
with $\phi$ being the phase of the coherent local oscillator.
In employing the BHD, we have assumed that the quantum state of the optical field can be accessed without inducing any disturbance.

We rewrite this to account for the two-level system scenario that we are considering by substituting the operators $\hat a, \hat a^\dagger $ with the $2\times2$ counterparts $\hat{\sigma}^-,\hat{\sigma}^+$, respectively. The corresponding PVM $\{\hat{\Pi}_{X_\phi}\}$ is given by the projectors on the two eigenstates of $\hat{X}_\phi$,

\begin{equation}\label{eq:BHD_pvm}
    \hat{\Pi}_{X_\phi}(k) = |e_k\rangle\langle e_k|, \quad k=1,2.
\end{equation}
\begin{figure*}[t]
    \centering
    \includegraphics[width=\linewidth]{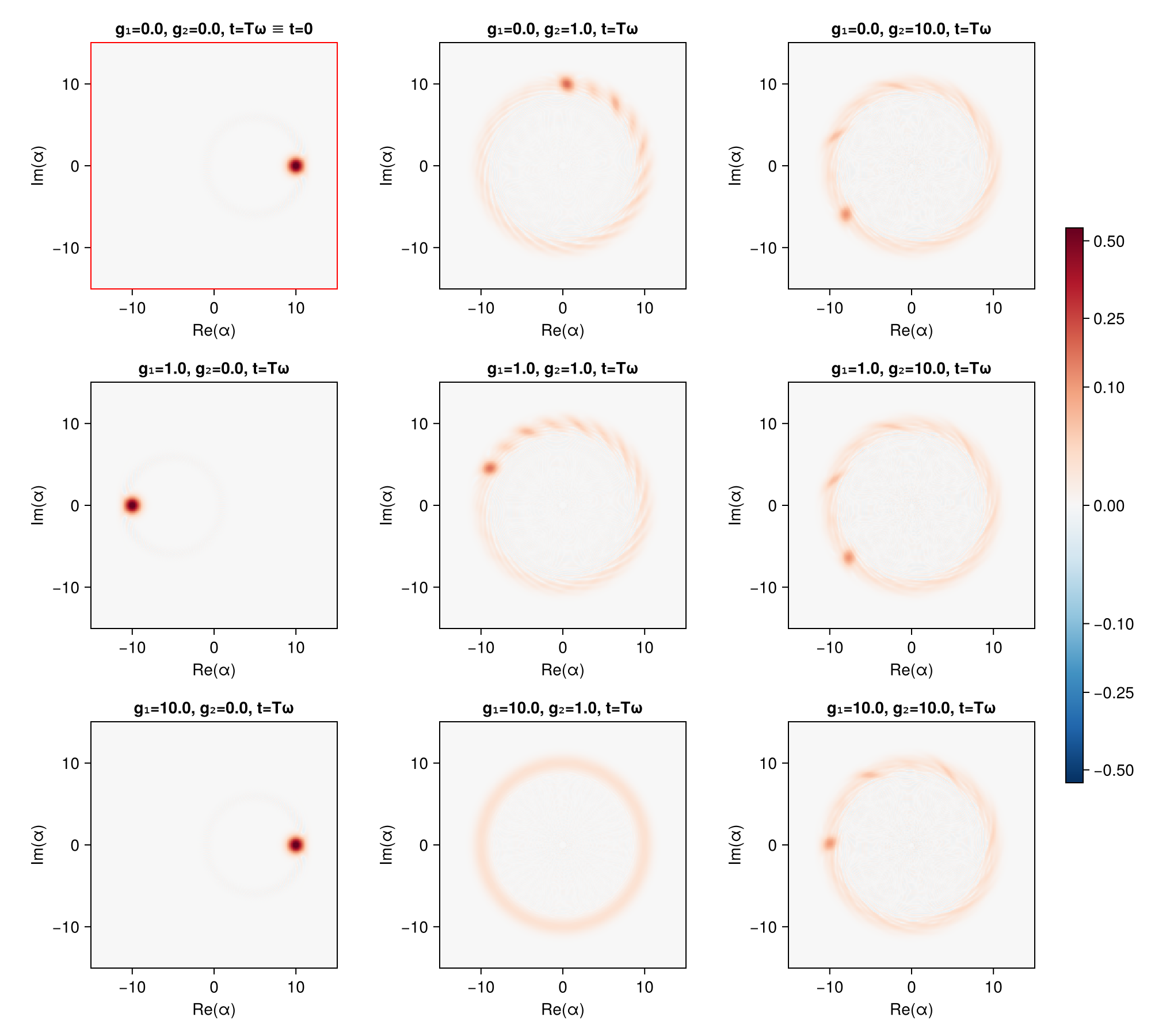}
    \caption{Wigner phase-space distribution of the optical field initially prepared in an coherent state, with $\alpha = 10$ and system parameters defined by Eq. \ref{eq:unit_free}. In the figures, $g_1$ takes the values from $0$, $1$ and $10$ left to right, while $g_2$ top to bottom. The plots are snapshots of the distribution taken at time equals $T_\omega$, defined in Eq. \eqref{eq:chtime}. With the border outlined in red, the top left figure is both a snapshot of the system at $T_{\omega}$ for $g_1= g_2 = 0$ but also the initial state for all the other figures at $t=0$.}
    \label{fig:wigner_coherent}
\end{figure*}
\section{Results}\label{sec:IV}

In this section, we address two main objectives. First, we present results on the system dynamics based on the general solution given in Eq. \eqref{eq:optical_coefficient_1}. As previously discussed, we assume the mechanical oscillator is initiated in a thermal state. For the optical field, the initial state is taken to be either
\begin{itemize}
    \item (i) a coherent state,
    \item (ii) quantum states within the subspace spanned by the vacuum and one-photon Fock states.
\end{itemize}
Due to the computational challenges associated with evaluating the symmetric logarithmic derivative (SLD), our discussion of the quantum Fisher information (QFI) and classical Fisher information (CFI) will be restricted to the $2\times 2$ photonic subspace, i.e, case (ii).

The results are organized into two main subsections. In subsection \ref{sec:independent}, we explore the dynamics of a dimensionless model in which the coupling parameters $g_1$ and $g_2$ are treated as independent variables. In subsection \ref{sec:interdependent}, we consider an experimentally relevant optomechanical setup, where $g_1$ and $g_2$ are interdependent due to a phenomenological model for $\omega(x)$.

\subsection{Independent couplings}\label{sec:independent}

Assuming that $g_1$ and $g_2$ are independent and can be varied individually, we investigate their respective and combined influences on the system dynamics. To facilitate a clear characterization of the system through dimensionless parameters, we adopt a unitless model defined by
\begin{equation}
\hbar = m = \Omega = n_{\text{th}} = \omega_c = 1. \label{eq:unit_free}
\end{equation}
Within this framework, we reveal an intrinsic periodicity in both the system dynamics and the associated QFI/CFI characteristics.
\begin{figure*}
    \centering    \includegraphics[width=\linewidth]{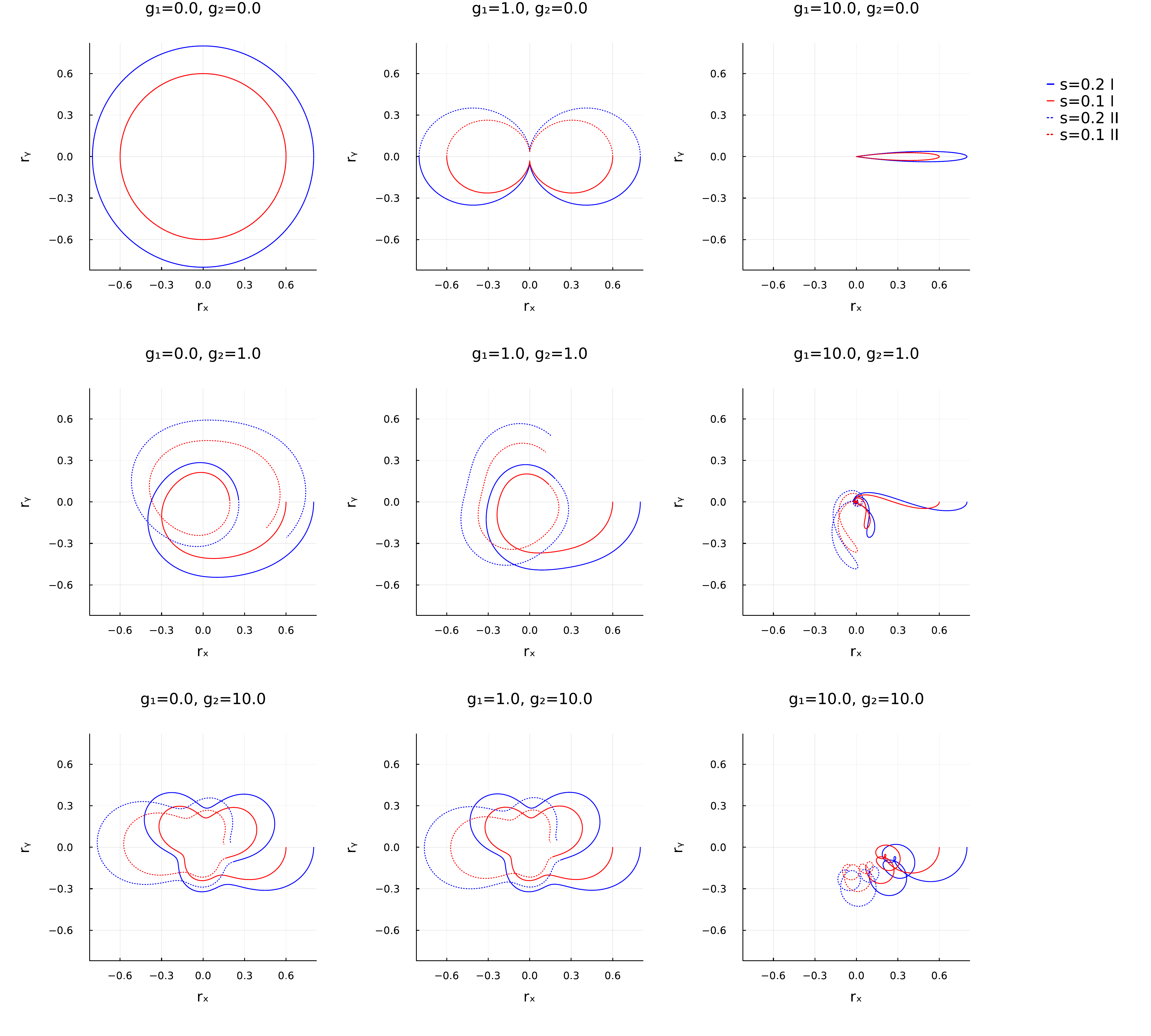}
    \caption{Trajectories in the X–Y plane of the Bloch sphere showing the evolution of the quantum states given in Eqs.~\eqref{eq:pure_state} and \eqref{eq:mixed_state}, parameterized by $s = 0.1$ (red) and $s = 0.2$ (blue). The system evolves from $t = 0$ to $t = 2T_\omega$, where $T_\omega$ is defined in Eq.~\eqref{eq:chtime}. The first cycle is shown with solid lines, and the second with dotted lines. The evolution depends solely on the parameter $s$ and is independent of the state’s purity. All other parameters are defined in Eq.~\ref{eq:unit_free}.}
    \label{fig:bloch_sphere}
\end{figure*}
\subsubsection{System dynamics for case (i)}

Under realistic conditions, the optical field is often initialized in a coherent states. Thus, we analyze the system dynamics using the Wigner phase-space distribution. This distribution provides a convenient and intuitive way to visualize the state of a quantized radiation field. It is closely related to the Glauber-Sudarshan $P$-representation \cite{mandel_optical_1995}, with the key difference lying in operator ordering: the $P$-representation corresponds to normal ordering of the operators $\hat{a}$ and $\hat{a}^\dagger$, while the Wigner phase-space distribution is associated with symmetric ordering. Given the density matrix of a field state $\hat{\rho}$, the Wigner phase-space distribution is given as follows:
\begin{equation}
W(\alpha,\alpha^*) =
\frac{1}{\pi^2} \int {\rm Tr}\left\{\hat{\rho} \, 
e^{\zeta \hat{a}^\dagger-\zeta^* \hat{a}}\right\}
 e^{\alpha\zeta^*-\alpha^*\zeta} d^2\zeta,
\label{Wignerf}
\end{equation}
where $\alpha,\zeta \in {\mathbb C}$.

In Fig. \ref{fig:wigner_coherent}, the Wigner phase-space distribution of the optical field is shown, with the field initially prepared in a coherent state of amplitude 
$\alpha=10$. The Fock space is truncated at $|\alpha|^2 + 4\sqrt{|\alpha|}$, corresponding to a maximum of 140 photons included in the calculations. The figure displays snapshots of the system’s Wigner phase-space distribution at time 
\begin{equation}
T_{\omega}= \frac{2 \pi}{\omega_1-\omega_0} \quad \text{and} \quad T_{\Omega}=\frac{2 \pi}{\Omega_1-\Omega_0} \label{eq:chtime}
\end{equation}
where $\omega_n$ is defined in Eq. \ref{eq:omn} and $\Omega_n$ in Eq. \eqref{eq:Omn}. These times correspond to intermediate characteristic timescales of the system’s evolution, considering the range of frequency differences among all values of $\omega_n$ and $\Omega_n$. In the figure, the coupling parameter $g_1$ increases from left to right, taking the values $0$, $1$, and $10$, while $g_2$ increases from top to bottom. In the Wigner phase-space representation, a coherent state appears as a localized Gaussian packet that undergoes circular motion around the origin in the absence of interaction between the optical field and the mechanical system. However, the interaction between the two oscillators causes the initially localized Gaussian packet to spread along the circular trajectory defined by the free evolution of the optical field. This occurs because the interaction Hamiltonian in Eq. \eqref{eq:Hamiltonian} commutes with $ \hbar \omega_c \hat{a}^\dagger \hat{a}$, i.e., with the free evolution of the optical mode. Within this circle, one can observe dense interference fringes. This behavior arises because a coherent state is a specific superposition of Fock states, whose probability amplitudes evolve according to Eq. \eqref{eq:optical_coefficient_1}, where $a_{n,m}=\alpha^n \alpha^{*m}/ \sqrt{n! m!} e^{-|\alpha|^2}$. Thus, the time-evolved state of the optical field is a general density matrix of size $140 \times 140$. Every density matrix is a convex combination of pure states, which are either Fock states or their superpositions. The convex combination of 
such states implies only a convex combination of their respective Wigner phase-space distributions. However, due to the large photon number involved up to $n=140$, pure states, unlike coherent states, induce a complex pattern of the interference fringes \cite{mandel_optical_1995}. The most interesting situation occurs when $g_2=0$, because then $T_\omega=2\pi/\omega_c$. In this situation, the first column of Fig. \ref{fig:wigner_coherent} shows the Wigner function of a coherent state, since the solution of Eq. \eqref{eq:optical_coefficient} reduces to: 
\begin{equation}
    a_{n,m}(t)=\frac{\alpha^n \alpha^{*m}}{\sqrt{n! m!}}e^{-|\alpha|^2} e^{-g_1^2 f^{(2)}_{n,m}(t)-f^{(0)}_{n,m}(t)}, 
\end{equation}
where
\begin{eqnarray}
f^{(0)}_{n,m}(t)&=i \omega_\text{c} t (n-m),\\
f^{(2)}_{n,m}(t)&=(2 n_{\text{th}}+1)\frac{1-\cos(\Omega t)}{\Omega^2} (n-m)^2\nonumber\\
&\qquad-i \frac{\Omega t-\sin(\Omega t)}{\Omega^2} (n^2-m^2).
\end{eqnarray}
Since all parameters in this section are set according to Eq. \eqref{eq:unit_free}, we have
\begin{eqnarray}
f^{(0)}_{n,m}(T_\omega)&=2 \pi i (n-m),\\
f^{(2)}_{n,m}(T_\omega)&=-2 \pi i (n^2-m^2).
\end{eqnarray}
Furthermore, since $n-m \in \mathbb{Z}$ and $g_1$ is chosen as $0$, $1$, and $10$, it follows that $g^2_1 (n^2-m^2) \in \mathbb{Z}$. Hence,
\begin{eqnarray}
    a_{n,m}(T_\omega)&=&\frac{\alpha^n \alpha^{*m}}{\sqrt{n! m!}}e^{-|\alpha|^2} e^{-g_1^2 f^{(2)}_{n,m}(T_\omega)-f^{(0)}_{n,m}(T_\omega)}, \nonumber \\
    &=& \frac{\alpha^n \alpha^{*m}}{\sqrt{n! m!}}e^{-|\alpha|^2}.
\end{eqnarray}
This explains why only a coherent state is observed in those plots.



\subsubsection{System dynamics for case (ii)}

To gain deeper insight into the system’s behavior, we consider case (ii), which permits an intuitive visualization of the dynamics on the Bloch sphere. Therefore, in this scenario, the initial photonic state $\hat \rho(0)$ is defined on the subspace spanned by the zero-photon state $|0\rangle$ and the one-photon state $|1\rangle$. Thus, $\hat \rho(0)$ is a $2\times 2$ positive semi-definite density matrix that can be written as

\begin{equation}\label{eq:pauli_expansion}
\hat \rho(0) = \frac{1}{2}(\hat{I}+r_x \hat \sigma_x+r_y\hat \sigma_y+r_z\hat\sigma_z),
\end{equation}

\noindent where $\hat \sigma_i$  with $i=x,y,z$ are the Pauli matrices and the coefficients $r_i$ are real numbers s.t. $r_x^2+r_y^2+r_z^2\leq 1$, with the equality satisfied when $\hat \rho(0)$ is a pure state.

In the following discussion, we consider either the pure state

\begin{equation}\label{eq:pure_state}
    |\psi_s\rangle = \sqrt{s}|0\rangle+\sqrt{1-s}|1\rangle,
\end{equation}

\noindent with $r_x(0) = 2\sqrt{s(1-s)}$, $r_y(0)=0$ and $r_z = (2s-1)$, or the mixed state defined as 

\begin{equation}\label{eq:mixed_state}
\rho_s = \frac{1}{2}|\psi_s\rangle\langle\psi_s|+\frac{1}{2}|\psi_{1-s}\rangle\langle\psi_{1-s}|,
\end{equation}
\noindent both parametrized by the single parameter $s$.
Note that the coefficients $r_x$ and $r_y$ of the pure states $|\psi_s\rangle$ and $|\psi_{1-s}\rangle$ coincide. 

In Fig.~\ref{fig:bloch_sphere} we show the parametric evolution of the optical field state~\eqref{eq:optical_coefficient} expressed in terms of the coefficients $r_x$ and $r_y$, allowing the system evolve between $t=0$ and $t=2 T_\omega$, defined in Eq. \eqref{eq:chtime}. The evolution is shown for two choices of the parameter $s$ of Eqs.~\eqref{eq:pure_state} and \eqref{eq:mixed_state}. In fact, by solving the integral in Eq.~\eqref{eq:optical_coefficient}, we can see that the diagonal elements of the optical field's density matrix are time-independent and therefore $r_z$ is a constant. 
Thus, the projection on the X-Y plane of the Bloch sphere suffices to describe the evolution of the optical system, while its value is independent from the purity $\mathcal{P}$ of the system, and can be parametrized solely by $s$. The complex behavior observed in the plots arises for the following reasons. After solving the unitary evolution of the entire system, we traced out the mechanical degrees of freedom without making any approximations. Consequently, the dynamics of the optical field alone reflects the situation in which the two oscillators become entangled and disentangled during the interaction. These processes, in turn, lead to a corresponding decrease or increase in the optical field’s purity. One observes the inward spiral of the coordinates towards the center of the Bloch sphere, which represents the maximally mixed state. Thus, the inward spiral shows a decrease in purity. This happens to a greater degree within the reference time $T_\omega$ for larger values of the coupling parameters, as expected from the increased interaction between the optical field and mechanical oscillator.

\begin{figure*}
    \centering
    \includegraphics[width=\linewidth]{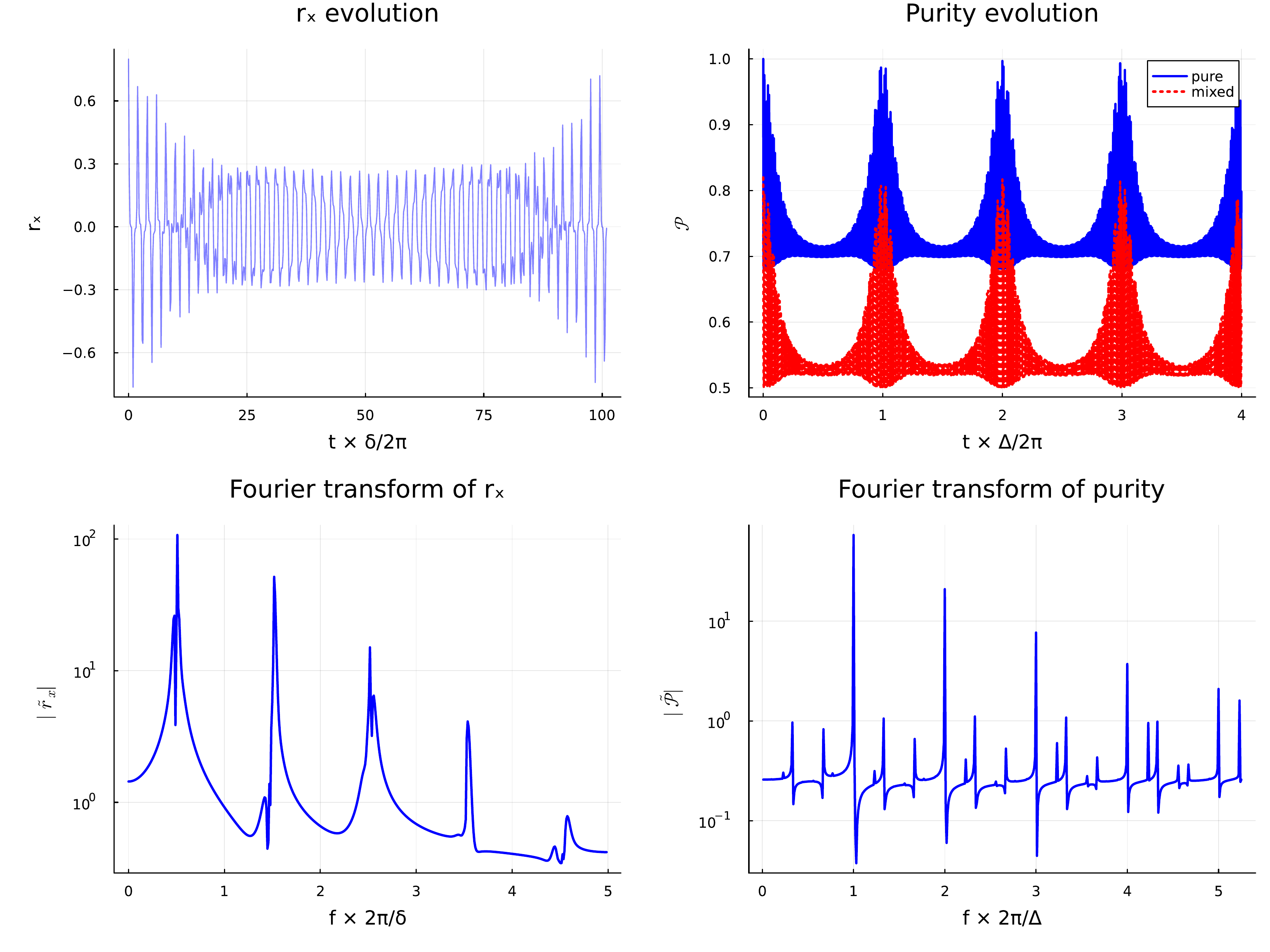}
    \caption{Time evolution of $r_x$, the $x$-component of the system on the Bloch sphere, plotted for the initial pure state $|\psi_s\rangle$ with $s = 0.2$. Time is expressed in units of the fast oscillation period, two full cycles of which are shown in Fig.~\ref{fig:bloch_sphere}. The fast oscillation is modulated by a slower beating pattern at frequency $\Delta = \Omega_1 - \Omega_0$. In all subplots, $g_1 = 1.0$ and $g_2 = 0.01$, while the remaining parameters are defined in Eq.~\ref{eq:unit_free}.}
    \label{fig:period}
\end{figure*}

\subsubsection{Spectral Analysis}

To demonstrate the system’s approximate periodic behavior, we examine the case where $g_2$ is small compared to $g_1$, i.e., $g_1 = 1$ and $g_2 = 0.01$. Note that this scenario, where $g_1$ is larger than $g_2$, is common in experimental setups, as we will present in Sec. \ref{sec:interdependent}. For this purpose, we show in Fig.~\ref{fig:period} the evolution of the $r_x$ component. After about $50$ fast cycles, the system spirals into the innermost point corresponding to the lowest purity, and then spirals outward again until it is back at the initial, outermost coordinates corresponding to the highest purity. This process happens at approximately the slower oscillation reference period $T_\Omega$, defined in Eq. \eqref{eq:chtime}. One can identify the relevant periods from the spectral analysis of $r_x$ and purity.
We plot the time evolution of the purity $\mathcal{P}$ in Fig.~\ref{fig:period} for a pure and mixed state with $s=0.2$.


In order to reveal the underlying frequencies, we performed a Fourier Transform on the dynamics of $r_x$ and $\mathcal{P}$, i.e.,


\begin{eqnarray}\label{eq:FFT_purity}
    \tilde{r}_x(f) &=& \int_{\mathbb{R}} dt \ e^{-ift}r_x(t), \\
    \tilde{\mathcal{P}}&=&\int_{\mathbb{R}} dt \ e^{-ift} \mathcal{P}(t).
\end{eqnarray}
Upon a closer examination of Fig.~\ref{fig:period}, we find that the primary frequencies of the $r_x$ oscillations occur at integer multiples of $\delta/2$, where $\delta = \omega_1 - \omega_0$. This behavior is also evident in Fig.~\ref{fig:wigner_coherent}, where, in the subplot corresponding to $g_1 = 1$ and $g_2 = 0$, the system completes exactly half a full cycle at $t = T_{\omega} = 2\pi / \delta$. While the oscillations and frequencies of $r_x$ take place on a shorter timescale, the most prominent periodicity of the system dynamics emerges on a longer timescale, $T_{\Omega}$, defined in Eq.~\eqref{eq:chtime} and illustrated in the right-hand subplots of Fig.~\ref{fig:period}. For both pure and mixed states, the system’s purity oscillates with an almost exact period of $T_{\Omega}$, as further confirmed by the Fourier transform of the purity, which exhibits clear peaks at integer multiples of $\Delta$, where $\Delta = \Omega_1 - \Omega_0$.

Mathematically, the explanation of such periodic behavior of the system can be found in the Gaussian integral that is used to calculate the time dynamics, whose exponent has both real and imaginary components. As the real components may cause damping of the system, the imaginary components cause oscillatory behavior instead. Among all the terms in the integrand exponent, 
 the two frequencies with larger amplitude are $\Delta=\Omega_1-\Omega_0$, see Eq. \eqref{eq:Omn}, and $\delta = \omega_1-\omega_0$, see Eq. \eqref{eq:omn}. Successive frequencies have an exponentially decaying amplitude and only appear in the spectral analysis as secondary peaks, as shown in Fig. \ref{fig:period}.



\subsubsection{Quantum and Classical Fisher Information}


In this subsection, we first present, using the setup from case (ii), a comparison between the CFI and QFI for various combinations of $g_1$ and $g_2$. Our goal is to demonstrate the saturation of the QFI by the CFI. We then examine the QFI properties for case (i). A more detailed analysis is provided in Sec.~\ref{sec:interdependent}, where the parameters of the model are assigned experimentally relevant values.
\begin{figure}[H]
    \centering
    \includegraphics[width=\linewidth]{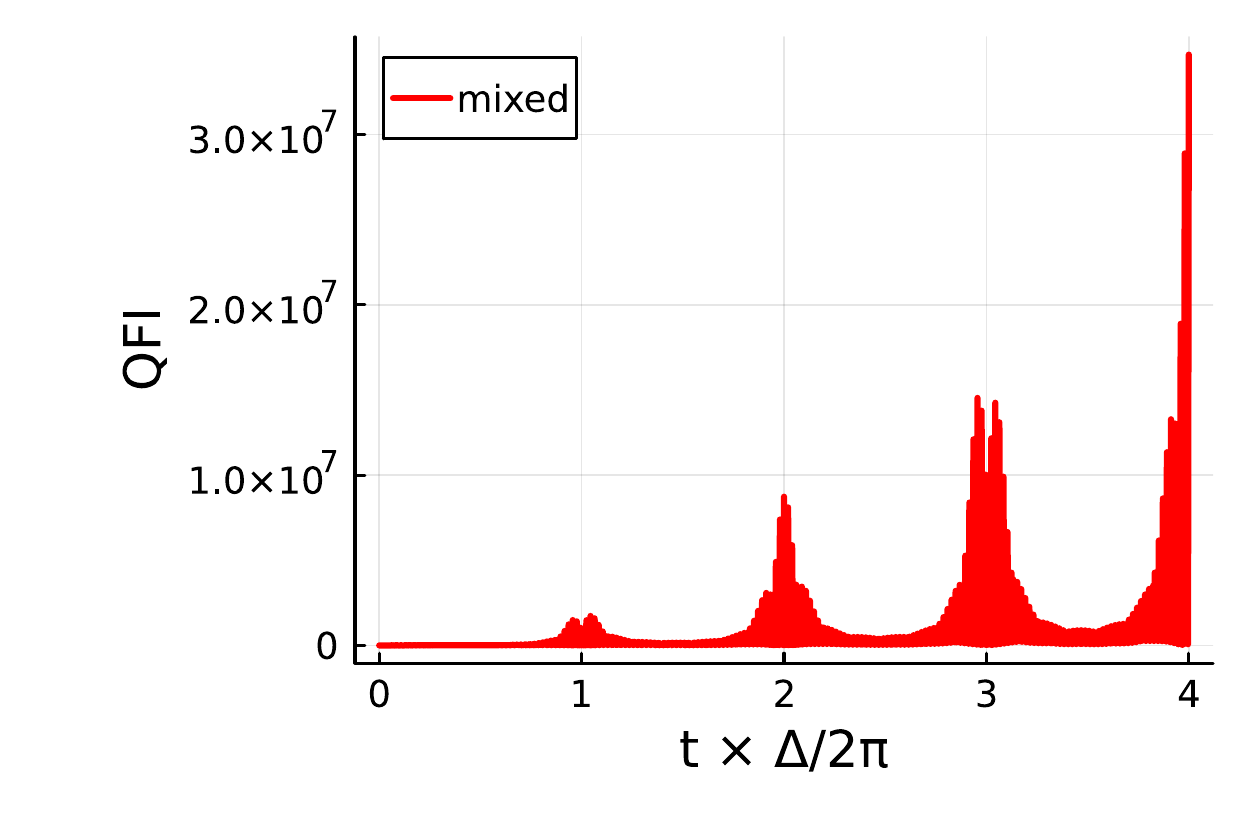}
    \caption{Quantum Fisher information as a function of time for the mixed state with $s = 0.2$, computed analytically. Periodic peaks appear at regular intervals of $2\pi/\Delta$, with $\Delta = \Omega_1 - \Omega_0$. The parameters are set as $g_1 = 1$, $g_2 = 0.01$, with all other relevant parameters set to unity, as in Eq.~\ref{eq:unit_free}.}
    \label{fig:QFI}
\end{figure}
In Fig.~\ref{fig:QFI}, we show the temporal evolution of the QFI for the initial mixed state~\eqref{eq:mixed_state} with $s=0.2$. We calculated the QFI using the simplified analytic expression for $\hat L_g$ given in Equation~\eqref{eq:2x2_QFI}.  Even for an initial mixed state, the periodicity of the peaks of the QFI coincide with the periodicity of the purity of the system. Around the peaks, the off-diagonal elements of the density matrix show greater sensitivity to changes of the $g_2$ parameter, as can be verified by solving Eq.~\eqref{eq:optical_coefficient}. We also observe that under the same parameter setup, the temporal evolution of the QFI for an initially pure state display much narrower and taller peaks, centered at the same time points as the mixes state counterpart, though this result is not shown here.

\begin{figure*}    \includegraphics[width=\linewidth]{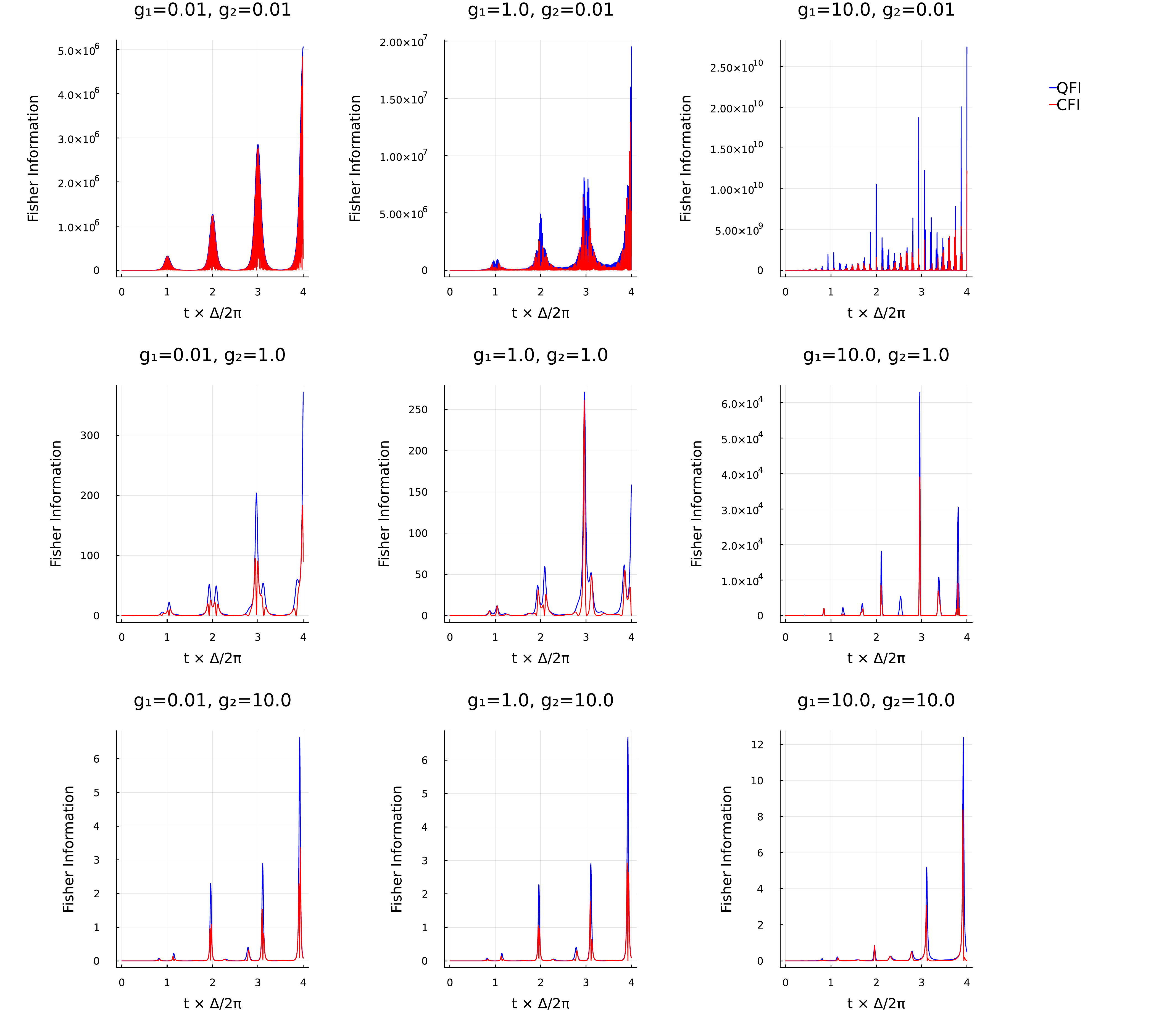}
    \caption{Comparison of quantum Fisher information (blue) and classical Fisher information (red) as a function of time for an initially mixed two-level optical field state given by Eq.~\ref{eq:mixed_state} with $s = 0.1$ and for $\Delta=\Omega_1-\Omega_0$. The time interval spans from $0$ to $4T_\Omega$, where $T_\Omega$ is defined in Eq.~\eqref{eq:chtime}. All other system parameters are set to unity, as in Eq.~\ref{eq:unit_free}.}
    \label{fig:Q_and_C_all}
\end{figure*}

In Fig. \ref{fig:Q_and_C_all}, various representative combinations of values for $g_1$ and $g_2$ are taken and the corresponding QFI and CFI evolution in time juxtaposed. In line with the method of BHD outlined in Sec. \ref{sec:III}, we choose the two-level system in case (ii) to calculate the CFI values for the system. The system for all subplots is initiated in a mixed state labelled by $s = 0.1$, see expression \ref{eq:mixed_state}. As time evolves from $0$ to $4T_\Omega$, one observes peaks at every regular interval of $T_\Omega$. The peaks also increase in intensity as time progresses and decreases in intensity as the values of the coupling parameters increase. Another feature of the correlation is that the peaks of QFI and CFI coincide.

\begin{figure}[H]
    \centering
    \includegraphics[width=\linewidth]{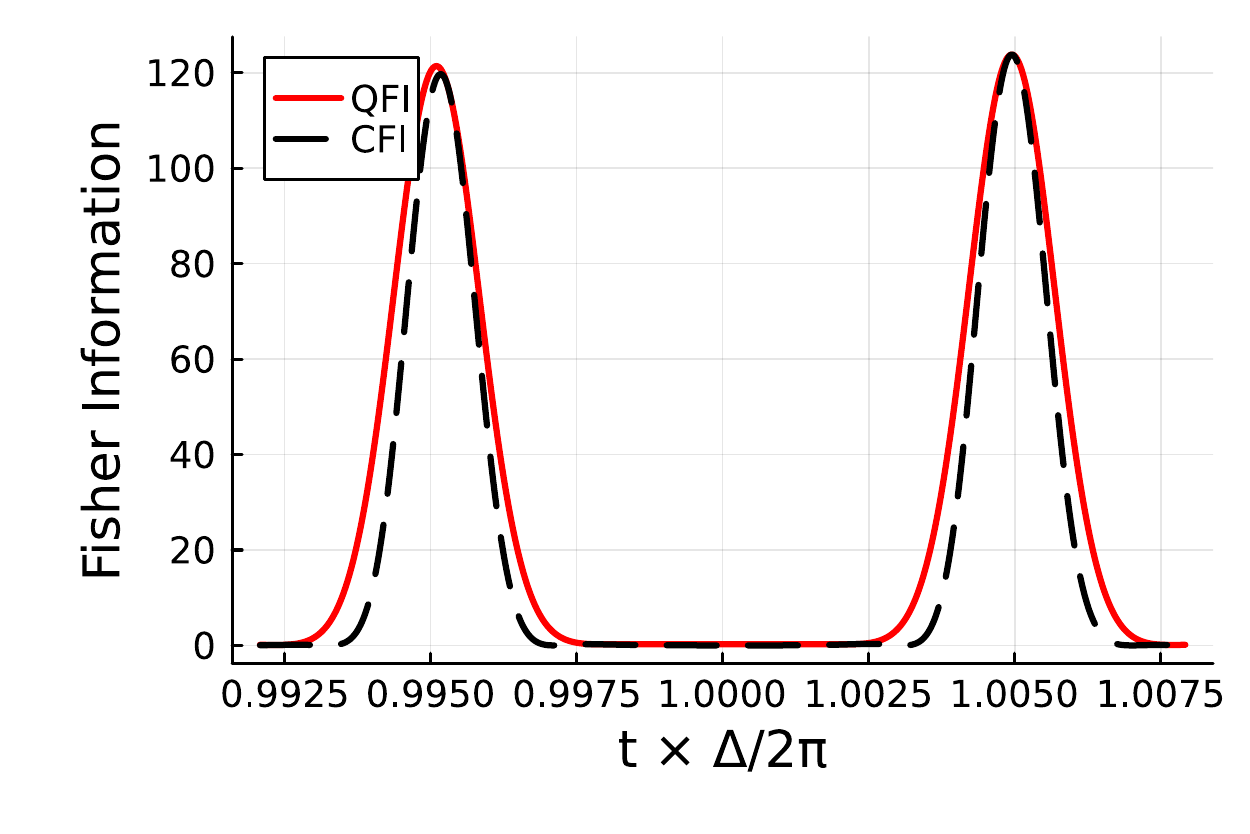}
    \caption{Classical Fisher information (black dashed line) and quantum Fisher information (red solid line) as a function of time, zoomed in around the first period of the system’s beating oscillation, $t = T_\Omega$, where $T_\Omega$ is defined in Eq.~\eqref{eq:chtime}. In this plot, $g_1 = 1.0$, $g_2 = 0.01$, $\phi = \pi/2$, and all other system parameters are set to unity, as in Eq.~\ref{eq:unit_free}. }
    \label{fig:q_vs_c_unpure}
\end{figure}

In Fig. \ref{fig:q_vs_c_unpure} we investigate the overlapping peaks of QFI and CFI closely to demonstrate the saturation behavior. By solving Eq.~\eqref{eq:optical_coefficient}, we see that the diagonal elements remain time-independent throughout the system evolution. The set of PVM related to the BHD, as defined in Eq.~\eqref{eq:BHD_pvm}, is chosen to calculate CFI, as it provides information about the off-diagonal elements. When the phase $\phi=0$ they reduce to
\begin{equation}
\Pi_0 = \vert +\rangle \langle + \vert,\; \Pi_1 = \vert -\rangle \langle - \vert,    
\end{equation}
where $\vert \pm\rangle = \frac{1}{\sqrt{2}}(\vert 0\rangle \pm \vert 1\rangle)$, which lead to the measurement probabilities of detecting one or no photon:
\begin{equation}
P_0 = \frac{1}{2}(1 + \rho_{01} + \rho_{10}), \; P_1 =   \frac{1}{2}(1 - \rho_{01} - \rho_{10}).    
\end{equation}
Note that $ \frac{1}{2}(\rho_{01} + \rho_{10})$ is simply the real part of both the off-diagonal elements. We compare the CFI obtained with the above described set of PVM to the QFI in Fig. \ref{fig:q_vs_c_unpure}, where we show the result around the slow oscillation reference time $t\approx T_\Omega$ in the case of an initial mixed state with $s=0.2$. The CFI has its peak is slightly shifted with respect to the QFI peak. However, we find that when the CFI is maximum, its value corresponds to the QFI value. Similar behavior is obtained both for different choices of $s$ and for an initial pure state.

\begin{figure}[H]
    \centering
    \includegraphics[width=\linewidth]{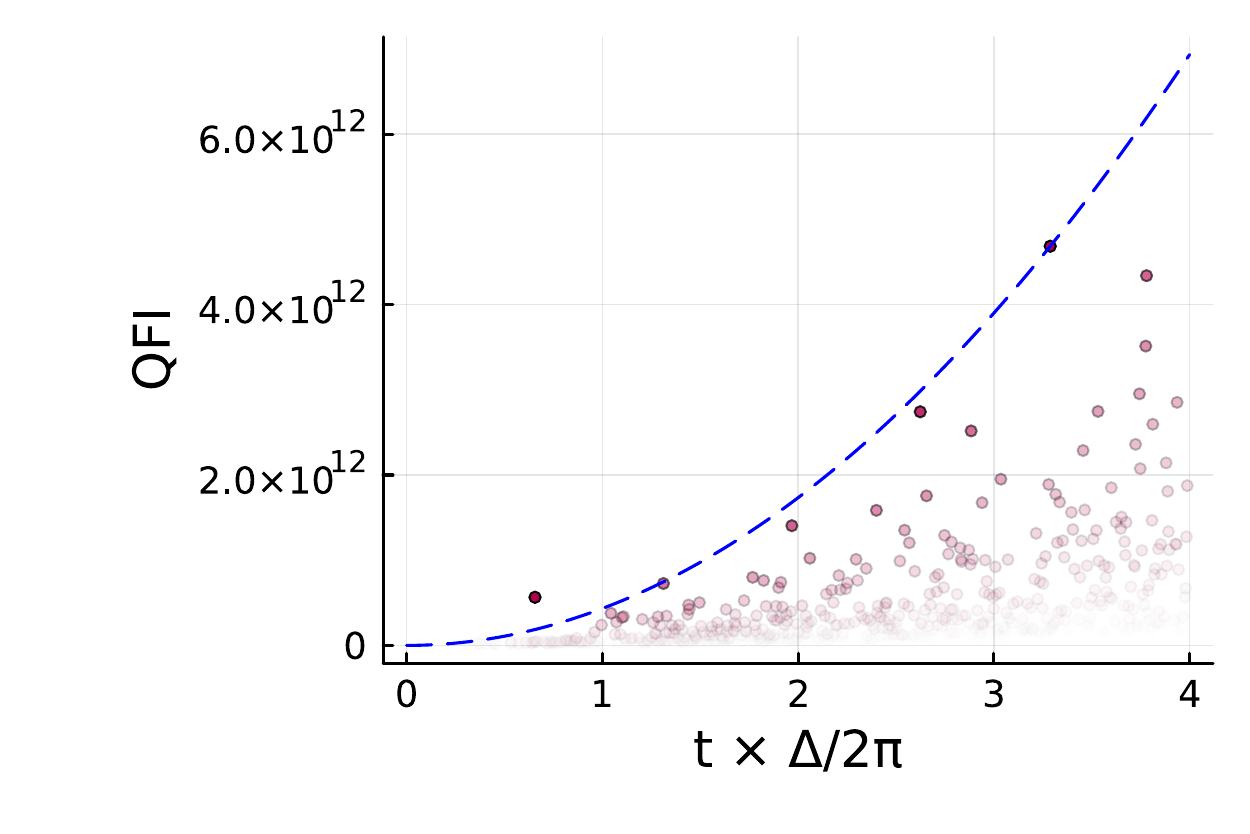}
    \caption{Quantum Fisher information against time as a scatter plot with $\Delta=\Omega_1-\Omega_0$. One observes peaks at regular interval with increasing magnitude. The system is initialized in a coherent state with $\alpha = 3$, with all system parameters set to unity as in Eq.~\ref{eq:unit_free}, while $g_1 = 1.0$ and $g_2 = 0.01$. The transparency of the data series is scaled by expression \eqref{eq:scaling_relation}. A quadratic relation over the peaks is plotted for reference with $Q \approx 4.334 \times 10^{11} \times  \frac{t^2}{T_\Omega^2}$. }
    \label{fig:QFI_coherent}
\end{figure}

Finally, we turn to case (i) and demonstrate the periodic behavior of QFI for an optical system initialized in a coherent state with $\alpha = 3$. As the system size has increased, the analytical two-level system solution to QFI no longer applies, hence numerical integration method is adopted. QFI as a function of time is presented as a scatter plot, see Fig. \ref{fig:QFI_coherent}. The transparency of the data series' coloring is taken as a linear function calculated as the following:
\begin{equation}
    \alpha = \frac{Q \times T^2_\Omega}{t^2 \times \max(\frac{Q \times T^2_\Omega}{t^2})}, \label{eq:scaling_relation}
\end{equation}
such that the transparency parameter $\alpha$ is a distribution on the interval $[0,1].$

Furthermore, a quadratic data series is juxtaposed on the scatter plot to fit the successive peaks observed, it follows the equation:
$$
y = 4.334 \times 10^{11} \times  \frac{t^2}{T_\Omega^2}.$$

From this, one sees that the intensity of the peaks scales quadratically with system time. This can be explained by the increasing amount of interaction between the optical field and the mechanical oscillator. Given that up to quadratic interaction is considered, a quadratic increase in the extractable information can be expected.

\subsection{Interdependent couplings}\label{sec:interdependent}

In the preceding sections, $g_1$ and $g_2$ were treated as independent variables. In a laboratory setting, however, certain phenomenological models approximate the cavity frequency as a function of some of the system parameters. 
In the following subsection, we show that such a model implies a mathematical relation between $g_1$ and $g_2$ when Eq.~\eqref{eq:expansion} is truncated strictly at the quadratic term; that is, all terms of order higher than two are assumed to have zero coefficients in the Taylor expansion. For the results presented below, we adopt this strict approximation and derive the corresponding relation between $g_1$ and $g_2$. 

We consider the following phenomenological model \cite{thompson_strong_2008}:
\begin{equation}\label{eq:omega_gs}
    \omega(x) = \frac{c}{L}\arccos\bigg{[}r\cos\frac{4\pi x}{\lambda} \bigg{]},
\end{equation}
where $L$ denotes the cavity length, $r$ the membrane reflectivity, $\lambda$ the wavelength of the laser driving the cavity, and $x$ is the
membrane center-of-mass position along the cavity axis \cite{CKLaw, karuza_tunable_2012}.
By combining Eq.~\eqref{eq:expansion} with Eq.~\eqref{eq:omega_gs}, an interdependent relation between $g_1$ and $g_2$ can be established. More specifically,
\begin{equation}
\omega'(x) =  \frac{cr}{L} \frac{\frac{4\pi}{\lambda}\sin\left(\frac{4\pi x}{\lambda}\right)}{\sqrt{1 - \left[r\cos\left(\frac{4\pi x}{\lambda}\right)\right]^2}},
\end{equation}
which has zeros at $x = (n\lambda)/2$, where $n$ is an integer. From Eq. \ref{eq:expansion}, we obtain
\begin{eqnarray}
\omega(x) &\approx& \omega(x_0) +\omega'(x_0)(x-x_0)+\frac{\omega''(x_0)}{2!}(x-x_0)^2 \nonumber \\
&=&\left[\omega(x_0) + \frac{\omega''(x_0)}{2!}x_0^2\right]  + \left[\omega'(x_0) - \omega''(x_0)x_0\right]x \nonumber \\
&+&\frac{\omega''(x_0)}{2!}x^2 =\omega_c +g_1 x + g_2 x^2,
\end{eqnarray}
To have $\omega'(x_0) = 0$, we set $x_0 = (n\lambda)/2$ with $n>0$, which yields the following relations
\begin{eqnarray}
\omega_c &=& \omega(x_0) + \frac{\omega''(x_0)}{2!}x_0^2, \label{eq:u1}\\
g_1&=&- \omega''(x_0)x_0, \label{eq:u2}\\
g_2 &=& \frac{\omega''(x_0)}{2!}. \label{eq:u3}
\label{eq:derive_g1}
\end{eqnarray}
\begin{table}[]
    \centering
    \begin{tabular}{|l|c r|}
        \hline
        & \textbf{Reference Values} &\\
        \hline
        $\hbar$  & $1.054 \times 10^{-34}$&$ [\text{J}\cdot\text{s}]$ \\
        \hline
        $m$  & $50 \times 10^{-15}$&$[\text{kg}]$ \\
        \hline
        $\Omega$  & $2\pi \times 134 \times 10^3$&$[\text{Hz}]$ \\
        \hline
        $n_{th}$ & $4.665 \times 10^7$ &\\
        \hline
        $\omega_c$ & $6.544 \times 10^{15}$&$[\text{Hz}]$ \\
        \hline
        $g_1$ & $- 1.308 \times 10^{18} $&$[\text{Hz}/\text{m}]$ \\
        \hline
        $g_2$& $1.445 \times 10^{23}$ & $[\text{Hz}/\text{m}^2]$ \\
        \hline
    \end{tabular}
    \caption{
    Table of parameter values used as a reference for an experimental system. All parameter values are expressed in SI units. Among the listed parameters, $m$ and $\Omega$ are directly taken from \cite{thompson_strong_2008}, while $n_{\text{th}}$, the average phonon number, is calculated for a temperature of $300,\text{K}$. The remaining parameters are obtained by substituting the experimental values of Ref. \cite{thompson_strong_2008}  required for the phenomenological model in Eq.~\eqref{eq:omega_gs} into Eqs.~\eqref{eq:u1}, \eqref{eq:u2}, and \eqref{eq:u3}.}
    \label{tab:param_standard}
\end{table}

For the results presented in this subsection, we use the numerical values of the system parameters reported in \cite{thompson_strong_2008}. The setup consists of a suspended $\text{SiN}$ membrane oscillating at a frequency of $\Omega = 2\pi \times 134\, \text{kHz}$, positioned at the center of a Fabry-P\'erot optical cavity of length $L = 67 \text{mm}$, and driven by an external laser with wavelength $\lambda = 1064\, \text{nm}$. This configuration yields an effective harmonic oscillator mass of approximately $m \approx 50\, \text{pg}$. The membrane has a reflectivity of $r = 0.42$, and the system operates at a temperature of $300\,\text{K}$. Under these conditions, the initial thermal state of the mechanical oscillator described in Eq.~\eqref{eq:initial_mechanical_state} corresponds to an average phonon number of $n_{\text{th}} \approx 4.665 \times 10^7$. This setup yields a relatively large quadratic optomechanical coupling of $g_2 \approx 1.445 \times 10^{23}\,\text{Hz}\cdot\text{m}^{-2}$, as obtained from Eq.~\eqref{eq:u3} when the membrane is positioned at either a node or an antinode of the optical field, i.e., $x_0 = n\lambda/2$, with $n = 400$. The remaining parameters are determined using Eqs.~\eqref{eq:u1} and \eqref{eq:u2} and are summarized in Table~\ref{tab:param_standard}. Furthermore, Refs.~\cite{sankey_strong_2010,karuza_tunable_2012} show that tilting the membrane by a small angle can induce coupling between different transverse modes of the electromagnetic field, thereby enhancing the quadratic coupling constant to $g_2 \approx 2.802 \times 10^{25}\,\text{Hz}\cdot\text{m}^{-2}$. 



\subsubsection{Estimates of the quadratic optomechanical coupling}

We aim to estimate the value of the optomechanical coupling constant, denoted by $g$, in an experimental context. To treat this as an unknown parameter, we assume that the phenomenological model given in Eq.~\eqref{eq:omega_gs} provides a good, though approximate, description of the coupling. To investigate this scenario, we use the parameter values listed in Table~\ref{tab:param_standard}. Note that throughout this subsection, we exclusively consider the mixed initial state corresponding to case (ii) in Section~\ref{sec:independent}.

\begin{figure}[H]
    \centering
    \includegraphics[width=\linewidth]{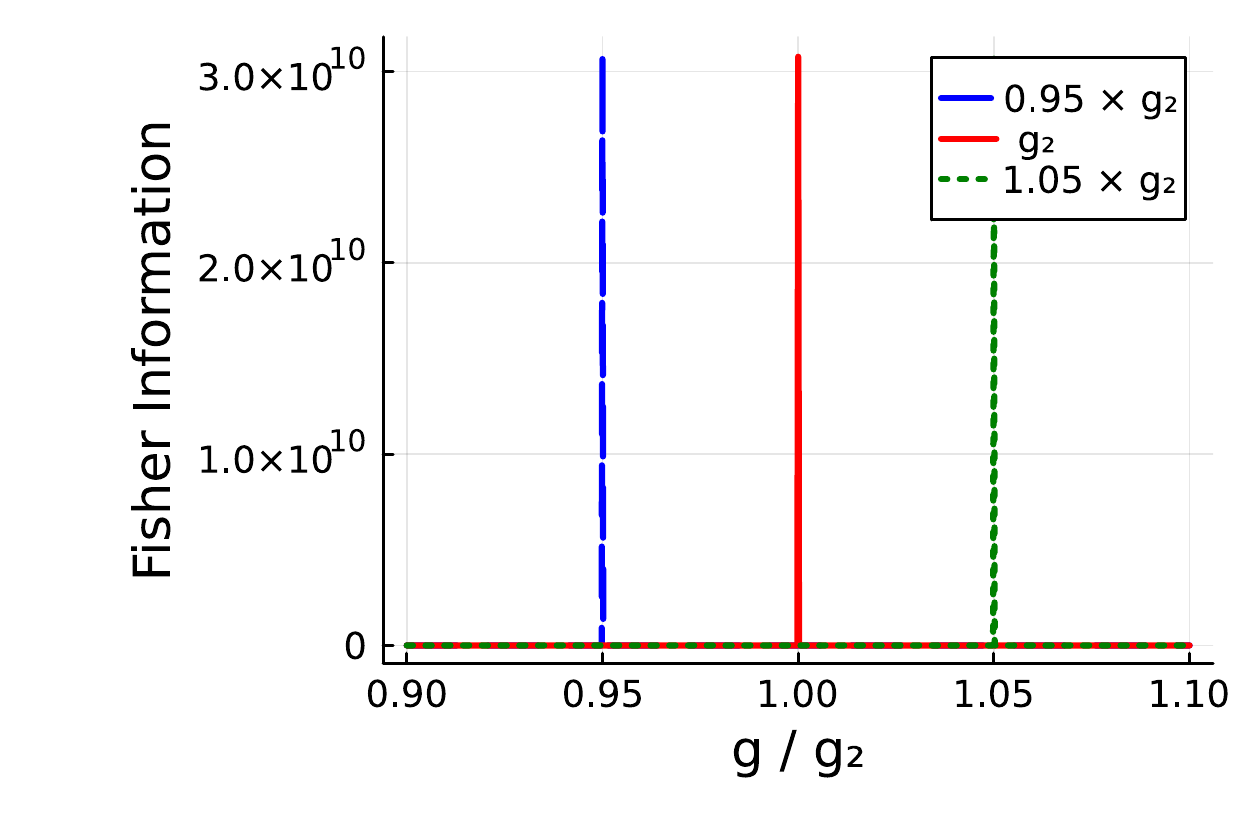}
    \caption{Quantum Fisher information of an experimental system as a function of the phenomenological values of $g_2$, evaluated at three different time points: $t = T_{g,\Omega} = \frac{2\pi}{\Omega_{1,g} - \Omega_{0,g}}$ for $g = 0.95g_2$ (blue dashed line), $g = g_2$ (red solid line), and $g = 1.05g_2$ (green dotted line). The range of the unknown coupling values $g$ considered spans from $0.9g_2$ to $1.1g_2$. All other system parameters are kept fixed at the values listed in Table~\ref{tab:param_standard}.}
    \label{fig: diff_g2}
\end{figure}

We compute and plot the QFI over a range of values of the quadratic coupling parameter $g$, centered around the phenomenological reference value $g_2$. For convenience, we introduce the notation $\Omega_{n,g} = \sqrt{\Omega^2 + \frac{2\hbar g_2 n}{m}}$ to emphasize the dependence of $\Omega_n$ on $g$ in Eq.~\eqref{eq:Omn}. Fig.~\ref{fig: diff_g2} presents three data series, each showing the QFI evaluated for different values of $g$ (on the $x$-axis) and at the reference peaking time associated with three specific choices of $g$, namely $g = 0.95g_2$, $g = g_2$, and $g = 1.05g_2$. The corresponding times are given by $t = T_{\Omega} = \frac{2\pi}{\Omega_{1,g} - \Omega_{0,g}}$.

The results show that, for each selected system time, the QFI exhibits a distinct sharp peak precisely centered at the corresponding $g$ value. This indicates a strong correlation between the system’s information-theoretic properties and the characteristic reference time $T_{\Omega}$ in a simple two-level system composed of zero- and one-photon Fock states. Furthermore, by comparing with Fig.~\ref{fig:QFI_coherent}, we observe that as more Fock states participate in the system dynamics, the periodicity of the QFI decreases, and the peaks no longer appear at integer multiples of $T_{\Omega}$.

\begin{figure}[H]
    \centering
    \includegraphics[width=\linewidth]{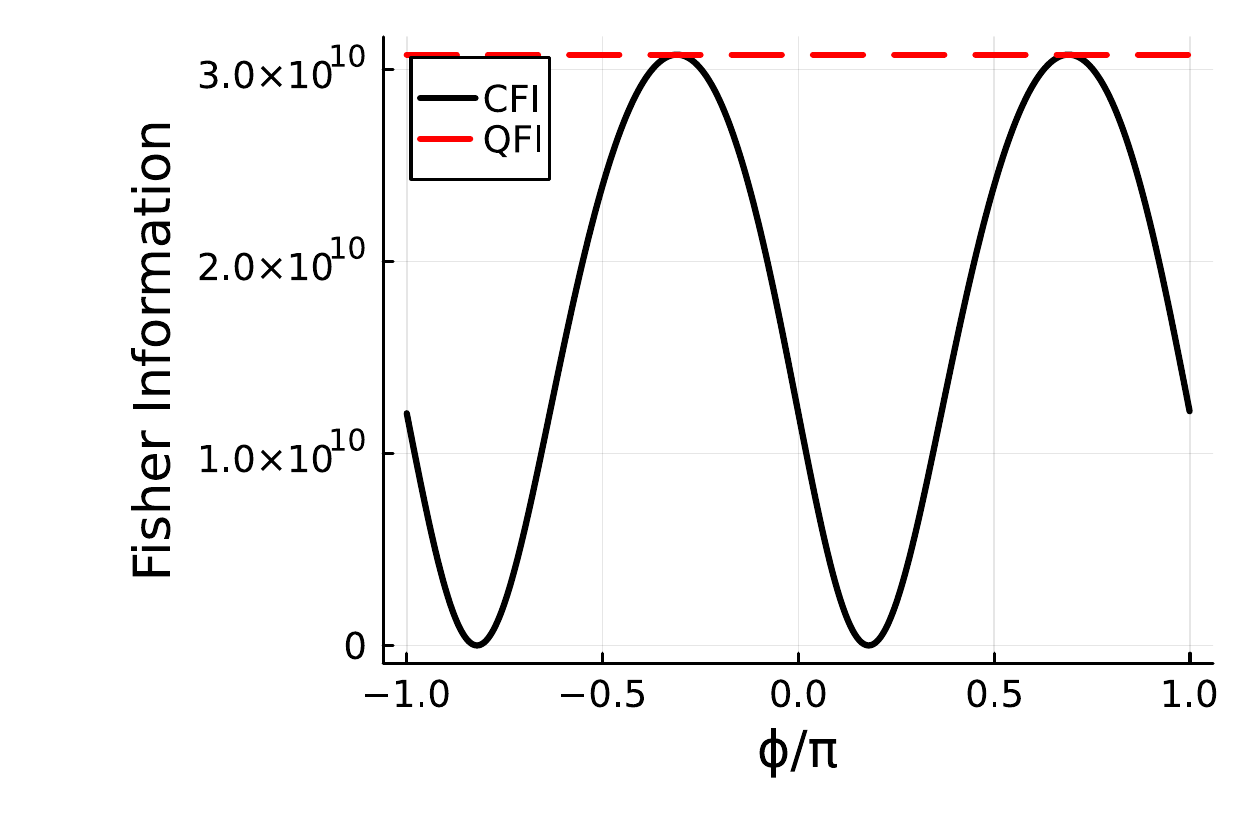}
    \caption{Quantum Fisher information (red dashed line) and classical Fisher information (black continuous line) for the experimental system, calculated at $t = 2T_\Omega$ as a function of the phase $\phi$. System parameters take on the values in Table \ref{tab:param_standard}.}
    \label{fig:phi_diff}
\end{figure}

Using these values, we found that the first peak of the QFI is $\approx 5.5\times 10^{15}$ for an initial mixed state $\hat \rho_s$ with $s=0.2$. We analyze how the choice of the parameter $\phi$ influences the CFI. While the behavior shown in Fig.~\ref{fig:q_vs_c_unpure} is reflected even with the experimental values, we see that a different choice of $\phi$ can lead the CFI to saturate the QFI. In Fig.~\ref{fig:phi_diff} we plot at $t=2 T_\Omega$, the CFI as a function of the phase $\phi$.

\begin{figure}[H]
    \centering
    \includegraphics[width=\linewidth]{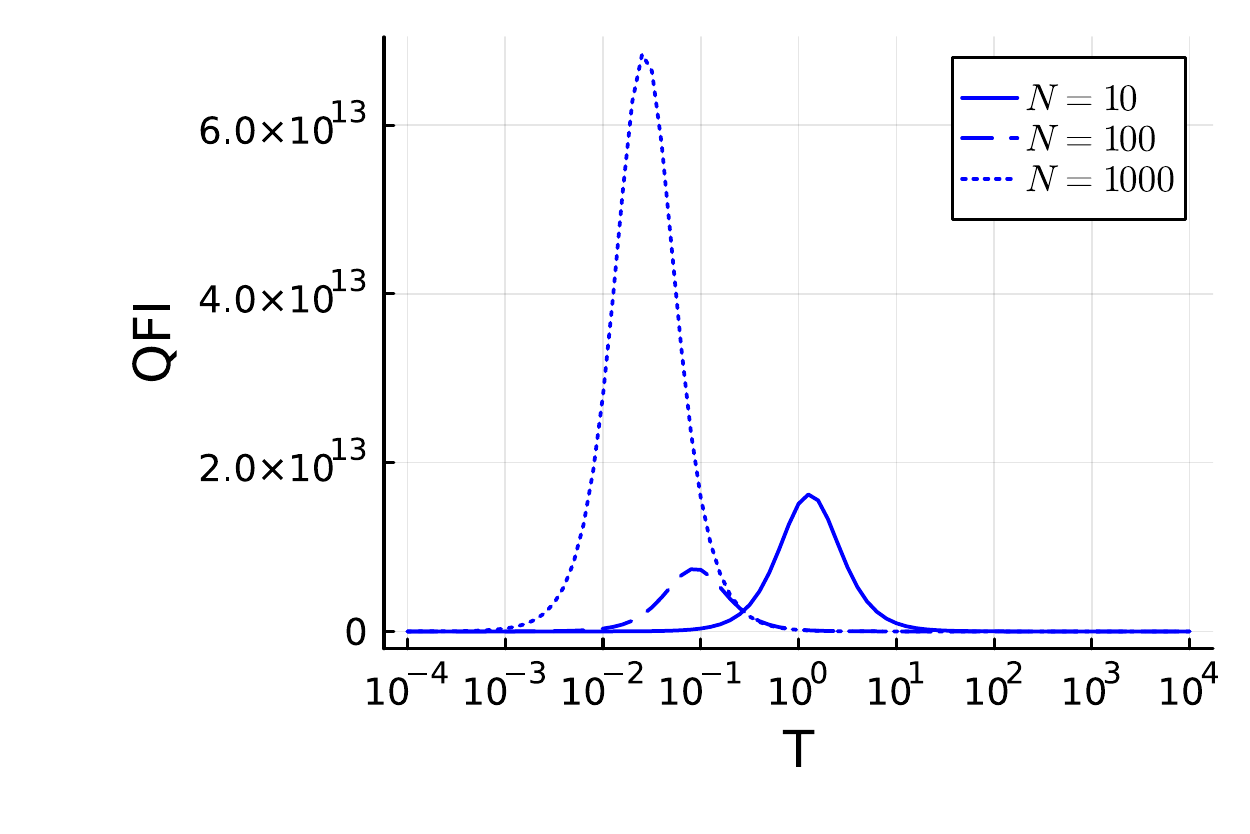}
    \caption{Quantum Fisher information as a function of the system temperature plotted at $t=N \times T_\Omega$, with $N=10$, $100$, and $1000$. Depending on the temperature the largest peak of the QFI is located at different periods of oscillation. The other system parameters are given in Table \ref{tab:param_standard}.}
    \label{fig:QFI_versus_T}
\end{figure}

Finally, we investigate the influence of temperature on the QFI. In Fig.~\ref{fig:QFI_versus_T}
we plot the QFI as a function of the temperature for the different peaks at $t=N \times T_\Omega$ with $N=10$, $10^3$, and $10^5$. When the temperature is close to zero the state of the mechanical system is substantially the ground state of the mechanical oscillator. This means that the optical system is interacting with a very weak field and consequently the information about the optomechanical coupling is low. 
At the opposite, when the temperature is large, the mechanical system is close to the maximally mixed state, which is persistent, regardless of the value of $g$. Even in this case, the information retrieval about $g$ is small. 
In between, there is a temperature where the QFI is at its maximum. 

From Fig.~\ref{fig:QFI_versus_T} we see that the largest peak of the QFI is located at a different oscillation period that depends on the temperature of the mechanical system. This behavior is opposite for cold and hot temperatures. 
For low $T$ although the mechanical system is approximately in the ground state, the information grows at each cycle, as the information available to the optical system cumulates. For high temperature $T,$ the mechanical bath tends to destroy all the available information and the peaks of the QFI decline at each cycle.

\section{Conclusion and perspectives}\label{sec:V}

In this work, we have investigated the dynamics of an optomechanical system with linear and quadratic couplings. We have found that each photon number contributes independently to the total Hamiltonian, which allows us to isolate these contributions and solve the temporal evolution of the density operator. Assuming that the optical and mechanical quantum states are initially factorized, we have derived a complete description of the optical field’s density operator using the formalism of two-photon coherent states. Furthermore, by considering the mechanical system to be initially in a thermal state, we obtained a solution that can be evaluated numerically.

We have investigated the dynamics of the system for two types of initial optical field states: a coherent state, and a state restricted to the subspace spanned by the zero- and one-photon Fock states. To understand the influence of the linear and quadratic optomechanical couplings on the system’s evolution, we first considered a dimensionless model in which the two couplings are treated as independent.

The evolution of the optical field has been analyzed using the Wigner phase-space distribution and, for the two-level subspace, the Bloch sphere representation. The Wigner distribution reveals a rich dynamics in which the photon number is preserved, while intricate interference fringes develop. In the two-level case, through a series of comprehensive analysis of the system dynamics that considers different combinations of $g_1$ and $g_2$, we find strong approximate periodicity in the system when one of the couplings is suppressed. As manifested in the experimentally relevant regime, where $g_1 \gg g_2$, the periodicity are attributed to the photon number-dependent frequencies of the mechanical oscillator.

To quantify the sensitivity of the optical field to variations in the quadratic coupling, we have calculated the quantum Fisher information for both initial conditions of the field's quantum state. The effect of the quadratic optomechanical coupling is particularly evident when the system regains its original purity, as the off-diagonal elements of the optical density matrix encode information about the coupling constant. Additionally, we have evaluated for the two-level case the classical Fisher information associated with balanced homodyne photodetection and found that it is capable of saturating the quantum Fisher information.

Finally, we have considered a phenomenological model for the cavity frequency, which results in interdependent linear and quadratic couplings. Using the experimental values from Eq.~\cite{thompson_strong_2008}, we have investigated the estimation of the quadratic coupling around the value predicted by the phenomenological model. Consistent with the findings from the dimensionless model, we have also demonstrated that in a balanced homodyne photodetection setup, adjusting the phase of the coherent local oscillator allows the classical Fisher information to saturate the quantum Fisher information.
Furthermore, we have studied the dependence of the quantum Fisher information to the temperature of the mechanical system, showing that it reaches its maximum at an intermediate temperature that allows the  state of the mechanical system to be highly populated while being far from the maximally mixed state.  

Our study describes the procedure to gain the most information on the optomechanical coupling constant out of the optical field inside the cavity. Future works need to assess whether our result remains valid when the measurement is performed on the output field, which can be obtained by filtering the input-output relation to the detector frequency~\cite{genes_robust_2008,sanavio_fisher-information-based_2020}.  

\begin{acknowledgments}
J.Z.B. acknowledges support from AIDAS-AI, Data Analytics
and Scalable Simulation, which is a Joint Virtual Laboratory gathering the Forschungszentrum Jülich and the French
Alternative Energies and Atomic Energy Commission, and from
the Hungarian National Research, Development and Innovation Office within the Quantum Information National Laboratory of Hungary (Grant No. 2022-2.1.1-NL-2022-00004).
\end{acknowledgments}

\appendix
\begin{widetext}
\section[\appendixname~\thesection]{Relevant properties of the two-phonon coherent state}\label{app:I}

In this appendix we write the overlap between the two-phonon coherent state $|\beta\rangle_n$ and the coherent state $|\alpha\rangle.$
It is~\cite{yuen_two-photon_1976}

\begin{eqnarray}
    \langle\alpha|\beta\rangle_n &=& \frac{1}{\sqrt{\mu_n}}e^{-\frac{1}{2}(|\alpha|^2+|\beta|^2+\frac{\nu_n}{\mu_n}\alpha^{*2}-\frac{\nu_n^*}{\mu_n}\beta^{2}-\frac{2}{\mu_n}\alpha^*\beta)}; \nonumber \\
    \label{eq:A1}
\end{eqnarray}
where the parameters $\mu_n,\nu_n$ are taken from the Bogoliubov transformations
\begin{eqnarray}
\hat{b}_n &=& (\mu_n\hat{b}+\nu_n \hat{b}^\dagger)\\
\hat{b}_n^{\dagger} &=& (\mu_n^*\hat{b}^{\dagger}+\nu_n^* \hat{b}).
\end{eqnarray}

\noindent and satisfy the relation

\begin{equation}
|\mu_n|^2-|\nu_n|^2=1.
\end{equation}

\noindent Their value is obtained by comparing the Bogoliubov equations~\eqref{eq:Bogoliubov}, with the above formula, yielding:

\begin{equation}
\mu_n = \frac{\Omega_n+\Omega}{2\sqrt{\Omega\Omega_n}},\quad\nu_n = \frac{\Omega_n-\Omega}{2\sqrt{\Omega\Omega_n}}.
\end{equation}
With the proper substitutions we can calculate all the implicit terms in Eq.~\eqref{eq:optical_coefficient}.

\section[\appendixname~\thesection]{Expressions for the multivariate Gaussian integral}\label{app:II}

In this appendix we explicitly write the elements of Eq.~\eqref{eq:optical_coefficient}. Consistently with the expression found in Eq.~\eqref{eq:optical_coefficient_1} and introducing the identity as in Eq.~\eqref{eq:identity_1}, we define the $\textbf{x}$ vector as
\begin{equation}
 \textbf{x} = \Big(\mathrm{Re}\{\alpha\},\mathrm{Im}\{\alpha\}, \mathrm{Re}\{\beta\},\mathrm{Im}\{\beta\},\mathrm{Re}\{\delta\},\mathrm{Im}\{\delta\},\mathrm{Re}\{\gamma\},\mathrm{Im}\{\gamma\}\Big), 
\end{equation}

\noindent and the matrix $A$ as
\[
A \;=\; 2\,I_{8} \;+\; M,
\]
with $M$ given by
\[
\begin{pmatrix}
\dfrac{2}{N_{th}} + F_+
& iF_-
& -\frac{1}{\mu_n} & \frac{i}{\mu_n} & 0 & 0
& -\frac{1}{\mu_m} & -\frac{i}{\mu_m} \\

iF_-
& \dfrac{2}{N_{th}} - F_+
& -\tfrac{i}{\mu_n} & -\tfrac{1}{\mu_n} & 0 & 0
& \tfrac{i}{\mu_m} & -\tfrac{1}{\mu_m} \\

-\tfrac{1}{\mu_n} & -\tfrac{i}{\mu_n}
& - \dfrac{\nu_n\bigl(1+e^{-2i\Omega_n t}\bigr)}{\mu_n}
& i\,\dfrac{\nu_n\bigl(1-e^{-2i\Omega_n t}\bigr)}{\mu_n}
& -\tfrac{e^{-i\Omega_n t}}{\mu_n}
& \tfrac{ie^{-i\Omega_n t}}{\mu_n}
& 0 & 0 \\

\tfrac{i}{\mu_n} & -\tfrac{1}{\mu_n}
& i\,\dfrac{\nu_n\bigl(1-e^{-2i\Omega_n t}\bigr)}{\mu_n}
& \dfrac{\nu_n\bigl(1+e^{-2i\Omega_n t}\bigr)}{\mu_n}
& -\tfrac{ie^{-i\Omega_n t}}{\mu_n}
& -\tfrac{e^{-i\Omega_n t}}{\mu_n}
& 0 & 0 \\

0 & 0
& -\tfrac{e^{-i\Omega_n t}}{\mu_n}
& -\tfrac{ie^{-i\Omega_n t}}{\mu_n}
& F_+
& -iF_-
& -\tfrac{e^{\,i\Omega_m t}}{\mu_m}
& \tfrac{ie^{\,i\Omega_m t}}{\mu_m} \\

0 & 0
& \tfrac{ie^{-i\Omega_n t}}{\mu_n}
& -\tfrac{e^{-i\Omega_n t}}{\mu_n}
& -iF_-
& - F_+
& -\tfrac{ie^{\,i\Omega_m t}}{\mu_m}
& -\tfrac{e^{\,i\Omega_m t}}{\mu_m} \\

-\tfrac{1}{\mu_m} & \tfrac{i}{\mu_m}
& 0 & 0
& -\tfrac{e^{\,i\Omega_m t}}{\mu_m}
& -\tfrac{ie^{\,i\Omega_m t}}{\mu_m}
& - \dfrac{\nu_m\bigl(1+e^{2i\Omega_m t}\bigr)}{\mu_m}
& i\,\dfrac{\nu_m\bigl(e^{2i\Omega_m t}-1\bigr) }{\mu_m}\\

-\tfrac{i}{\mu_m} & -\tfrac{1}{\mu_m}
& 0 & 0
& \tfrac{ie^{\,i\Omega_m t}}{\mu_m}
& -\tfrac{e^{\,i\Omega_m t}}{\mu_m}
& i\,\dfrac{\nu_m\bigl(e^{2i\Omega_m t}-1\bigr)}{\mu_m}
& \dfrac{\nu_m\bigl(1+e^{2i\Omega_m t}\bigr)}{\mu_m}
\end{pmatrix},\]

where the following shorthands were used:
$$F_+ = \tfrac{\nu_n}{\mu_n} + \tfrac{\nu_m}{\mu_m}, \; F_- = \tfrac{\nu_n}{\mu_n} - \tfrac{\nu_m}{\mu_m}.$$
The vector $b$ has the form
\begin{equation}
b = -\frac{1}{2}
\begin{pmatrix}
0\\
0\\
2\Bigl\{\Re\left[\eta_n(t)\right]\cos(\Omega_n t)-\Im\left[\eta_n(t)\right]\sin(\Omega_n t)
-\eta_n(t)\,\tfrac{\nu_n}{\mu_n}e^{-i\Omega_n t}\Bigr\}
-\eta_n(t)\,e^{i\Omega_n t}
+\eta^*_n(t)\,e^{-i\Omega_n t}\\
2\Bigl\{\Re\left[\eta_n(t)\right]\sin(\Omega_n t)+\Im\left[\eta_n(t)\right]\cos(\Omega_n t)
-i\,\eta_n(t)\,\tfrac{\nu_n}{\mu_n}e^{-i\Omega_n t}\Bigr\}
+i\,\eta_n(t)\,e^{i\Omega_n t}
+i\,\eta^*_n(t)\,e^{-i\Omega_n t}\\
-\tfrac{2}{\mu_n}\,\eta_n(t)
-\tfrac{2}{\mu_m}\,\eta^*_m(t)\\
\tfrac{2 i}{\mu_n}\,\eta_n(t)
-\tfrac{2i}{\mu_m}\,\eta^*_m(t)\Bigr)\\
2\Bigl\{\Re\left[\eta_m(t)\right]\cos(\Omega_m t)-\Im\left[\eta_m(t)\right]\sin(\Omega_m t)
-\eta^*_m(t)\,\tfrac{\nu_m}{\mu_m}e^{i\Omega_m t}\Bigr\}
-\eta^*_m(t)\,e^{-i\Omega_m t}
+\eta_m(t)\,e^{i\Omega_m t}\\
2\Bigl\{\Re\left[\eta_m(t)\right]\sin(\Omega_m t)+\Im\left[\eta_m(t)\right]\cos(\Omega_m t)
+i\,\eta^*_m(t)\,\frac{\nu_m}{\mu_m}e^{i\Omega_m t}\Bigr\}
-i\,\beta^*_m(t)\,e^{-i\Omega_m t}
-i\,\eta_m(t)\,e^{i\Omega_m t}
\end{pmatrix}.
\end{equation}
Finally, the scalar $c$ has the following explicit form:
\[
c \;=\; -\tfrac12 \Bigl[\lvert \eta_n(t)\rvert^2 + \lvert \eta_m(t)\rvert^2
\;-\;\frac{\nu_m}{\mu_m}\,\eta^{*2}_m(t)
\;-\;\frac{\nu_n}{\mu_n}\,\eta_n^2(t)\Bigr].
\]
\end{widetext}
\bibliographystyle{apsrev4-1}
\bibliography{main}

\end{document}